\documentclass [11pt,letterpaper]{JHEP3}
\usepackage{epsfig,array,cite,latexsym,amsmath,oldgerm}
\usepackage{graphicx}
\usepackage{epsfig}
\usepackage{epic}       
\usepackage{cite}	
\usepackage{psfrag}     
\usepackage{bm}		
\usepackage{verbatim}   
\advance\parskip 1.0pt plus 1.0pt minus 2.0pt	
\def\Nc{N_{\rm c}}

\def\beq{\begin{eqnarray}}
\def\eeq{\end{eqnarray}}
\newcommand\bal{ \begin{align}}
\newcommand\eal{\end{align} }

\newcommand\eqn[1]{\label{eq:#1}} 
\newcommand\Eq[1]{Eq.~\eqref{eq:#1}}

\newcommand\bfz{\mathbf{Z}}
\newcommand\bfl{\boldsymbol{\Lambda}}

\newcommand{\CM}{{\cal M}}
\newcommand{\CN}{{\cal N}}
\newcommand{\CP}{{\cal P}}

\newcommand{\CQ}{{\cal Q}}

\newcommand{\bfn}{{\bf n}}

\DeclareMathOperator{\Tr}{Tr\,}

\newcommand{\sla}[1]%
        {\kern .25em\raise.18ex\hbox{$/$}\kern-.55em #1}

\newcommand{\mybar}[1]%
        {\kern 0.6pt\overline{\kern -0.6pt#1\kern -0.6pt}\kern 0.6pt}
\newcommand{\dig}{\kern-1.5pt \raisebox{.9ex}{$\cdot$}  \kern1.5pt
  \raisebox{0ex}{${\mathbf\cdot}$}\kern1.5pt \raisebox{-.9ex}{$\cdot$}} 
\newcommand{\digb}{\kern-1.5pt \raisebox{.75ex}{$\cdot$}  \kern1.5pt
  \raisebox{0ex}{${\mathbf\cdot}$}\kern1.5pt \raisebox{-.75ex}{$\cdot$}} 
\newcommand{\digc}{\kern-1.5pt \raisebox{1.05ex}{$\cdot$}  \kern1.5pt
  \raisebox{0ex}{${\mathbf\cdot}$}\kern1.5pt \raisebox{-1.05ex}{$\cdot$}} 
\preprint {BUHEP 05-12}
\title
    {%
Supersymmetric Deformations  of  Type IIB Matrix Model 
as Matrix Regularization of $\CN=4$ SYM
    }%
\author
    {%
     Mithat \"Unsal
    \\Department of Physics
    \\Boston University 
    \\590 Commonwealth Ave, Boston, MA 02215
    \\Email: 
    \parbox[t]{2in}{\email {unsal@buphy.bu.edu} }
    }%
\abstract
    {%
We construct a $\CQ=1$  supersymmetry and $U(1)^5$  global symmetry
preserving deformation of the type IIB matrix model.   This model, 
 without orbifold   projection,
serves as 
a nonperturbative regularization for 
$\CN=4$ supersymmetric  Yang-Mills theory  in four Euclidean 
dimensions. 
Upon deformation, the eigenvalues of the bosonic matrices are forced 
to reside  on the surface of a hypertorus. 
 We explicitly show the relation between    the  noncommutative moduli 
space of the deformed  matrix theory  and the Brillouin zone of the emergent 
lattice theory. This observation makes the transmutation of the moduli space  
into  the   base space of target  field
theory clearer. 
 The lattice theory is slightly nonlocal, however the 
nonlocality is suppressed by the lattice spacing. In the classical 
continuum limit, we recover the   $\CN=4$ SYM theory.
We also discuss the result in terms of  D-branes and interpret it as   
 collective excitations of D(-1) branes forming  D3 branes. 
  }%
\begin {document}
\setlength{\baselineskip}{1.25\baselineskip}
\section {Introduction}
\label{sec:1}
The study of supersymmetric deformations of the type IIB matrix model  
(or IKKT matrix model \cite{Ishibashi:1997xs}) is interesting both from 
the field theory and string
theory points of view.  The deformations may   alter   the classical 
moduli spaces  
into noncommutative spaces,   and hence, may form a bridge between 
zero dimensional,  finite rank  matrix theories and 
higher dimensional regularized quantum field theories.  
In this work, we show such a correspondence between  a $\CQ=1$   
supersymmetry and $U(1)^5$ global symmetry preserving deformation   
of type IIB matrix model, and $\CN=4$ supersymmetric Yang-Mills theory in four
dimensions.   
Mainly,   the deformed type IIB matrix model  serves as  finite $N_c$ 
matrix regularization  (which will be
shown to be equivalent to a lattice regularization), 
where $N_c$ is the number of colors,
 for $\CN=4$ supersymmetric Yang-Mills theory 
in four dimensions.  This construction can be thought as supersymmetric 
generalization of the 
relation between   twisted Eguchi-Kawai (TEK) and  
 four dimensional lattice gauge theory \cite{Gonzalez-Arroyo:1982hz,  
Ambjorn:1999ts}.

The essential point of the construction is the relation between  two
seemingly different concepts. The noncommutative moduli space of vacua, which
is the zero action configuration of the deformed bosonic action,  
and the noncommutative base space, the space that the target field theory
lives in. In the type IIB theory without deformation, the eigenvalues of the
bosonic matrices can be anywhere in the moduli space. 
 The deformations, that will be considered in this paper,    
alter the structure of moduli space in such a way that the 
eigenvalues of bosonic matrices are forced to reside on the surface of a 
four dimensional  hypertorus, $T^4$. This is essential for the 
emergence of four dimensional  spacetime and will be discussed in detail in 
sections \ref{sec:3} and \ref{sec:fluc}. 

In the description of the deformed type IIB action in section \ref{sec:2}, 
we use the notation 
introduced in    \cite{Kaplan:2005ta}, which makes the  
$SU(5) \times U(1)$ subgroup  of global $SO(10)$ R-symmetry and $\CQ=1$
supersymmetry  manifest.  
Therefore, 
 we use five complex matrices $z^m$ and their 
hermitian conjugates    $\mybar z_m$ as the bosonic degrees of freedom 
 instead of ten bosonic hermitian matrices.  This notation is more akin 
to our four dimensional habits and the bosonic action becomes just a  sum 
of D-term  and F-terms, which are derivatives of superpotential.  
The bosonic action of the deformed  type IIB matrix model is given by 
\beq  
S_{bosonic}&&=\frac{\Tr}{g^2} \bigg[ 
\frac{1}{2}
 \bigg( \sum_{m=1}^{5} [\mybar z_m, z^m] \bigg)^2 
  + \sum_{m\neq n =1}^{5}| e^{-i \Phi_{mn}/2} \mybar z_{m}
\mybar z_{n} - e^{i \Phi_{m n}/2} \; \mybar z_{n}
\mybar z_{m}  |^2 
\bigg]   
\label{eq:Sb}
\eeq
where  the antisymmetric phase matrix $\Phi_{mn}$ are the deformation 
parameters. This  change in  the action may be considered as the 
deformation of the ``superpotential''.  
 The  action  \Eq{Sb} may be obtained by the dimensional 
reduction of the four dimensional hypercubic lattice action of $\CN=4$ 
SYM  to zero volume (a point)  upon the  use of twisted boundary
conditions. 
 The  phases  $\Phi_{\mu \nu}$,  where $\mu, \nu=1,\ldots,4$, 
are associated with electric and 
magnetic background fluxes  \cite{'tHooft:1979uj} 
and    $\Phi_{\mu5}= - \sum_{\alpha=1}^{4} 
\Phi_{\mu\alpha}$. 
We will show that  a two parameter family of deformation which, for example, 
 corresponds 
to electric and magnetic fields in z-direction,
suffices to generate the target $\CN=4$ theory  
\footnote{This type of deformation 
appeared in the EK context under the rubric 
  ``twisting''  \cite{Gonzalez-Arroyo:1982hz,
Gonzalez-Arroyo:1983ac, Das:1984nb},  and in the context of
supersymmetric gauge theories in four dimensions as  a marginal deformation of
  superpotential (as $\beta$ deformation) 
of $\CN=4$ theory to $\CN=1$ (see for example \cite{leigh:1995ep, 
Berenstein:2000ux, Dorey:2003pp})} . 

The technique described here is different from the 
recent  constructions of  lattice 
regularization for  extended supersymmetric Yang-Mills
theories \cite{Kaplan:2002wv, Cohen:2003aa,Cohen:2003qw, Kaplan:2005ta} and 
deconstruction of six dimensional theories 
\cite{ArkaniHamed:2001ca, ArkaniHamed:2001ie}. 
The current
approach does not require orbifold projection of a mother theory.  As it will 
be shown explicitly,  the deformation of the  mother  theory  becomes 
 the lattice regularization. 
Recall that the orbifold projected matrices  are  sparse and encode a  
commutative spacetime lattice. In our approach the spacetime lattice 
is generated by the zero action configurations of the deformed type IIB    
matrix model  and there are no constraints on our matrix variables.
Similarly, based on  \cite{Kaplan:2002wv} and 
\cite{Ambjorn:2000cs,  Ambjorn:1999ts}, the lattice action 
for noncommutative counterparts of the extended  supersymmetric
gauge theories are also constructed, by using orbifolds with discrete
torsion,  in   \cite{Nishimura:2003tf} for spatial 
and \cite{Unsal:2004cf} for Euclidean two dimensional lattices. 
(For recent progress and independent approaches on supersymmetric gauge 
theories 
on lattice,  also see  \cite{  Catterall:2005fd,  
lattice}. For a  review on 
 discrete torsion, see \cite{Sharpe:2000ki} and the references in 
\cite{Unsal:2004cf})

To show the equivalence of the deformed type IIB matrix model to the 
four dimensional theory on a discrete torus, we decompose the bosonic 
matrices into 
a background and fluctuations around it.  
By using  the  background field techniques, it is possible to exhibit  that 
 the fluctuations around the zero action
  configurations  
can be  identified  as a  lattice  action 
on a four dimensional  discrete torus.  This  action has many 
parallels to   the lattice action in ref.\cite{Kaplan:2005ta}.  
The  main difference with respect to 
the action in  \cite{Kaplan:2005ta} is the noncommutativity of the base 
space lattice, which can be traded with the nonlocalness of the interactions.
However, 
we show that  two types of finite-volume continuum limits are
 possible. In a particular scaling of the deformation parameter (in which the
 noncommutativity length tends to zero with respect to the linear extent of
 the box), the  interaction localizes in the position space and 
the outcome  is 
ordinary local field theory. In the  limit where   the 
noncommutativity length  is nonvanishing, 
the outcome is nonlocal $\CN=4$ SYM.   

It is also possible to interpret the result of this paper within string
theory. The main result can be understood as a 
collective excitation of a
finite number  of D(-1)-branes to generate discretized  
D3-brane(s) wrapped on a four dimensional hypertorus.   
Notice that we do not  use 
  compactification  (see \cite{Taylor:1997dy} for a review) to generate 
the higher dimensional theory. Also, our  D(-1)-brane configurations  are 
inherently different from the  BPS-saturated backgrounds discussed in 
\cite{Aoki:1999vr}
, which have
typically finite action and are unstable at finite $N_c$. 
 We discuss this
direction briefly in  section \ref{sec:5}. 
The rest of the   paper is organized as follows: In section \ref{sec:2} 
and appendix 
\ref{sec:ap1}, 
we set  the notation and derive the deformed type IIB action.  In section  
\ref{sec:3}, we analyze the noncommutative moduli space and make the relation
with TEK more explicit. In section  \ref{sec:4}, we show that the  the
deformed type IIB action can be rewritten as a lattice action and obtain the 
classical continuum limit.

\section{The type IIB matrix model  and 
its supersymmetric deformations}
\label{sec:2}
The type IIB  matrix model  can be obtained by  the dimensional 
reduction of the ten   
dimensional $\CN=1$  $U(N_c)$ supersymmetric Yang-Mills  theory 
to zero  dimension. 
The theory possess a $G_R =SO(10)$  global R-symmetry, inherited 
by the (Euclidean) Lorentz symmetry of the ten dimensional theory.
The bosonic and fermionic degrees of freedom of the matrix theory are  
in vector  (${\bf 10}$) and positive chirality spinor (${\bf 16}$) 
representations   of  $SO(10)$.  
The action of the type IIB matrix model 
is conventionally expressed as 
\beq 
S=-\frac{1}{g^2} \sum_{\alpha \neq \beta=1}^{10} \frac{1}{4}
\Tr \ [v_\alpha, v_\beta]^2  + \frac {i} {2} \Tr \omega^T C \Gamma_{\alpha} 
 [v_\alpha, \omega]
\label{eq:typeIIB}
\eeq
where $v_\alpha=v_{\alpha}^a t^a$ is a Hermitian matrix 
and  $\omega= \omega^a t^a $ is a 32-component spinor  which has only 
its upper  sixteen component nonvanishing.  The ten gamma matrices 
$\Gamma_{\alpha}$ form a chiral basis of $SO(10)$ and  $C$ is charge
conjugation matrix.  
The Lie
algebra generators are normalized such that $\Tr t^a t^b = \delta^{ab}$. 
This action  respects  $\CQ=16$ supersymmetries generated   
by $\delta v_\alpha = - \kappa^T C \Gamma_{\alpha} \omega, \;\; {\rm and} \;\;
 \delta 
\omega =  \frac{i}{4} [v_{\alpha}, v_{\beta}]\; [  \Gamma_{\alpha},  
\Gamma_{\beta}] \kappa $ ,  
where $\kappa$ is a positive chirality Grassmann spinor.   

We aim to construct   deformations of the type IIB matrix model, which
preserves  $\CQ=1$ supersymmetry and $U(1)^5$  global symmetry.  Since the
bosonic  variables that appear  in the action  \Eq{typeIIB} 
are not eigenstates  of the $U(1)^5$, this form of the action is not adequate 
for our purpose.  The type IIB action, in terms of eigenstates of $SU(5)
\times U(1)$, which provides the right basis for our purpose,  
 is constructed in   \cite{Kaplan:2005ta} and is outlined in the next section.

The motivation behind this construction is 
to obtain a finite rank matrix regularization of $\CN=4$ SYM theory in four 
dimensions. It is well-known that the TEK model is a matrix regularization 
of (noncommutative) Yang-Mills theories in even dimensions.  
Our goal here is to construct a fully supersymmetric counterpart of this  
relation.  It is, therefore,  important to construct the supersymmetric 
matrix model analog of TEK. The TEK is obtained by dimensional reduction 
of a fully regulated lattice gauge theory to a single site by using the 
twisted boundary conditions. We will follow the same strategy and   
reduce the lattice regulated $\CN=4$ SYM to a single point. 

\subsection{$\CN=4$ SYM on hypercubic lattice}
\label{sec:lattice}
In order to construct the deformed Type IIB matrix model,  we start with the
four dimensional hypercubic orbifold lattice. In \cite{Kaplan:2005ta}, the 
lattice action of $\CN=4$ SYM is given for both $A_4^{*}$ lattice and 
hypercubic lattice.  Here, we will use hypercubic for convenience.   The
dimensional reduction of the lattice to a single site by using periodic
boundary conditions  reproduces the type IIB matrix model in the eigenbasis of 
$U(1)^5$ symmetry.   The reduction by using    twisted boundary
conditions gives the  deformed Type IIB  action.   

\setlength{\extrarowheight}{5pt}
\begin{table}[t]
\centerline{
\begin{tabular}
{|c|c|c|}
\hline
$p$-cell & fields  & orientations  
\\ \hline
0-cell & $ \lambda  $ & $ 0 $ 
\\ \hline
1-cell & $z^{\mu}, \psi^{\mu}; \qquad \mybar z_{\mu}  $ & $ {\bf \hat e_{\mu}}; 
\qquad {-\bf \hat e_{\mu}}     
  $ 
\\ \hline
2-cell & $ \xi_{\mu \nu}  $ & $  {-\bf \hat e_{\mu} - \hat e_{\nu}  }  $ 
\\ \hline
3-cell & $ \xi^{ \nu \rho \sigma}  $ & 
$ {\bf \hat e_{\nu}  + \hat e_{\rho} + 
\hat e_{\sigma}} = {-\bf \hat e_{\mu} - \hat e_{5}} $ 
\\ \hline
4-cell & 
$( z_{ \mu \nu \rho \sigma}, \psi_{ \mu \nu \rho \sigma})   ; 
\qquad 
\mybar z^{ \mu \nu \rho \sigma} $ & $ {\bf \hat e}_{5}= - \sum_{\mu=1}^{4}
{\bf \hat e}_{\mu} ; \qquad  - {\bf \hat e}_5
       $ 
\\ \hline
\end{tabular}
}
\caption{\sl The  distribution of the bosonic and fermionic fields 
to the $p$-cells of the hypercubic lattice.  Each field is totally
antisymmetric in its indices.  For convenience, we will also use 
$\xi^{ \nu \rho \sigma} = {\epsilon^{ \nu \rho \sigma \mu }} \xi_{\mu 5}$,
$( z_{ \mu \nu \rho \sigma}, \psi_{ \mu \nu \rho \sigma})=  
\epsilon_{\mu \nu \rho \sigma}(  z^5, \psi^5)$, 
and $\mybar  z^{ \mu \nu \rho \sigma}= \epsilon^{\mu \nu \rho \sigma}
\mybar z_5$.   The  cartesian basis vectors  are given by  
 $({\bf \hat e}_{\mu})_\nu = \delta_{\mu \nu}$ where $\mu, \nu = 1, \ldots 4$
\label{tab:tab1}.}
\end{table}

Let us first rewrite the action of the $\CN=4$ SYM theory in $d=4$ 
dimensional hypercubic lattice  and establish our  notation.   
The  fundamental cell  of a  hypercubic lattice 
contains one site, four links, six faces,  four cubes and one hypercube,   
collectively called  $p$-cells. 
There are bosonic  and fermionic fields   associated with each $p$-cell.   
For a $d=4$ dimensional lattice, the total number of $p$-cells  in a 
unit-cell is 
\beq
2^4 = \sum_{r=0}^{4}  \binom{4}{k} = 1 \oplus 4 \oplus 
6 \oplus 4 \oplus 1
\eeq
The fermions, which form the  ${\bf 16}$ dimensional spinor 
representation  of the $SO(10)$ are distributed among the $p$-cells. 
The labeling of  the fermions follows the  name of cell that it  
lives in:  $\omega \sim {\bf 16} = \lambda \oplus \psi^{\mu} \oplus 
\xi_{\mu\nu} \oplus \xi^{\mu\nu \rho} \oplus \psi_{\mu\nu \rho \sigma} = 
1 \oplus 4 \oplus 6 \oplus 4 \oplus 1   $
where each fermion is totally antisymmetric tensor and they reside on  
sites, links, diagonal of faces, diagonal of the cubes, and the body diagonal
of the hypercube \cite{Rabin:1981qj}
 The upper and lower indices determine the orientation of
the fermionic link field   
(see Table.\ref{tab:tab1}). 
For notational convenience and brevity,  we introduce 
$ \xi_{\mu 5} = \frac{\epsilon_{\mu \nu \rho \sigma}}{3!}  
\xi^{\nu \rho \sigma}  $ and 
$ \psi^{5} = \frac{\epsilon_{\mu \nu \rho \sigma}}{4!}  
\psi^{\mu \nu \rho \sigma}  $.\footnote{  
The idea of associating the $p$-cells with  $p$-form fermions  also 
appears as a natural implementation of Dirac-K\"ahler fermions on the lattice.
See for example \cite{Catterall:2005eh} and reference therein.}

The bosons, which form a ${\bf 10}$ dimensional vector representation of 
$SO(10)$ are associated with  1-cells and body diagonal of 4-cells. 
We have  
  $v_{\alpha} \sim {\bf 10} = z^{\mu} \oplus \mybar z_{\mu}  
 \oplus z_{\mu\nu \rho \sigma}  \oplus \mybar z^{\mu\nu \rho \sigma} =
 4 \oplus \mybar 4 \oplus 1 \oplus 1   $. For convenience, we rename 
$z^5= \frac{\epsilon^{\mu\nu \rho \sigma}}{4!}  z_{\mu\nu \rho \sigma}$ and 
$\mybar z_5= \frac{\epsilon_{\mu\nu \rho \sigma}}{4!}  
\mybar z^{\mu\nu \rho \sigma}$  \footnote{ Throughout this paper,  we adopt the
convention that repeated indices are summed, where $\alpha,\beta\ldots$
are $SO(10)$ indices summed over $1,\ldots,10$; the indices
$\mu,\nu,\dots$ are $SU(4)$ indices summed over $1,\ldots,4$;  
$m,n,\ldots$ are $SU(5)$ indices summed over $1,\ldots,5$.  
In the description of the continuum (which will be used only in section 4.2
and 4.3 )
the indices $a,b,\ldots$ are $SO(6)$ $R$-symmetry indices summed over 
$1,\ldots,6$; and  $\mu,\nu,\dots$ are $SO(4)$ spacetime indices summed over
$1,\ldots,4$. We hope the latter wont cause an inconvenience for the reader. 
}.   

Naturally, the $\CQ=1$ superfields on the hypercubic lattice are 
associated with  $p$-cells.   
The component form of the  superfields and 
the $p$-cell that they are residing  in are  given as
\begin{equation}
\begin{aligned}
&0-{\rm cell}: \qquad {\bf \Lambda}_{\bfn} = \lambda_{\bfn} -i\theta  d_{\bfn} \ ,\\
&1-{\rm cells}:  \qquad {\bfz}^{\mu}_{\bfn} = z^{\mu}_{\bfn} + \sqrt{2}\,\theta \,
  \psi^{\mu}_{\bfn}\ , \qquad  \mybar z_{\mu, \bfn} \\
 &2-{\rm cells}:\qquad {\bf \Xi}_{\mu \nu,\bfn}= \xi_{\mu \nu, \bfn} -  
2\theta\,\, \mybar E_{{\mu \nu},\bfn}
\\
&3-{\rm cells}: \qquad  {\bf \Xi}_{\mu 5,\bfn}= \xi_{\mu 5, \bfn} -  
2\theta\,\,  \mybar E_{{\mu 5},\bfn}  \\
&4-{\rm cell}: \qquad  {\bfz}^{5}_{\bfn} = z^{5}_{\bfn} + \sqrt{2}\,\theta \,
\psi^{5}_{\bfn}\ , \qquad   \mybar z_{5, \bfn}  \\
\end{aligned}
\eqn{superfields}
\end{equation}
where  the $\mybar z_{\mu, \bfn}$ and   $\mybar z_{5, \bfn}$  are 
supersymmetry singlets and the $d_\bfn$ is an auxiliary field.
 The antiholomorphic functions $\mybar E_{{\mu \nu},\bfn}$ and  
 $\mybar E_{{\mu \nu},\bfn}$ are given by 
\beq 
& \mybar E_{{\mu \nu},\bfn}=  
 \mybar z_{\mu,\bfn+ {\bf \hat e_{\nu}} }  \mybar z_{\nu,\bfn} -
 \mybar  z_{\nu,\bfn+ {\bf \hat e_{\mu}} }  \mybar
 z_{\mu, \bfn}  \cr
&  \mybar E_{{\mu 5},\bfn}= \mybar z_{\mu,\bfn+ {\bf \hat e_{5}} }  
\mybar z_{5,\bfn}-  
\mybar  z_{5,\bfn+ {\bf \hat e_{\mu}} }  \mybar z_{\mu, \bfn}
\eqn{E}
\eeq
The antiholomorphic $\mybar E$  functions are  supersymmetry singlets 
as well  since they only depend on 
$\mybar z$, but not $z$. They can be thought as zero 
dimensional counterpart of the superpotential (more precisely, the F-term)

The hypercubic lattice action may be written in 
manifestly  $\CQ=1$ supersymmetric form as in \cite{Kaplan:2005ta}
\footnote{\label{convention} 
The description of plaquette types and the flux passing through 
them  are easier to understand 
when we make the $SU(4)$ indices $\mu, \nu \ldots $ manifest, 
However, it is more economical to use the  
$SU(5)$ notation 
occasionally (by keeping in mind  the geometric structures that the fields 
are associated with). Hence, we will continue to use both. 
Hereafter, we sometimes  combine the $\mu\nu \rho \sigma$
 index which runs over $1,\ldots 4$   with $5$ and  use the Latin letters 
  $mnpqr$ which ranges  over  $1,\ldots 5$. Mainly, we
combine the four link multiplets and  one 4-cell multiplet as  
$\bfz^{\mu} + \bfz^5 
\rightarrow \bfz^{m}$ and  $\mybar z_{\mu} + \mybar z_5 \rightarrow \mybar
z_m$, 
the six fermi multiplets on the faces and four fermi multiplets on the 
3-cells combine into  ${\bf \Xi}_{\mu\nu} + {\bf \Xi}_{\mu 5} \rightarrow 
{\bf \Xi}_{mn}$ and naturally  $ \mybar  E_{\mu\nu} +  
\mybar E_{\mu 5}\rightarrow  \mybar E_{mn}$.}
\begin{eqnarray}
S &=& \frac{\Tr}{g^2} \sum_{\bf n} \bigg[ \int \;  d \theta \left( 
-\frac{1}{2} {\bf \Lambda_{\bf n}} {\partial}_{\theta} {\bf \Lambda_{\bf n}}  
- {\bf\Lambda_{\bf n}}(\mybar z_{m,\bf {n - \hat e_m}} 
{\bf Z}_{m,\bf {n - \hat e_m}}
 - {\bf Z}_{m, \bf n} \mybar z_{m, \bf n} )  + 
\frac{ 1}{2}
{\bf \Xi}_{mn, \bf n}{\bf E}_{mn, \bf n}
\right) \cr \bigg.
&+& \bigg.
  \frac{\sqrt 2 }{8}  \epsilon^{mnpqr} {\bf \Xi}_{mn, \bf n}
(\mybar z_{p, \bf {n-  \hat e_p}}  {\bf \Xi}_{qr, \bf {n +  \hat e_m + 
\hat e_n }} - 
{\bf \Xi}_{qr, \bf {n -  \hat e_q - \hat e_r }} \mybar z_{p, \bf {n +  
\hat e_m + \hat e_n }}) \bigg] \qquad \qquad
\eqn{lattice}
\end{eqnarray}
where the holomorphic ${\bf E}$ functions are  
\beq 
{\bf E}_{mn, {\bf n}}= ({\bf Z}_{m,\bf n} {\bf Z}_{n, \bf {n + \hat e_m}}- 
{\bf Z}_{n, \bf n} 
 {\bf Z}_{m, \bf {n + \hat e_n}})\; .
\eeq
The ${\bf E}$ function is a  function 
of  superfields $\bfz$ and has a  $\theta$ 
expansion where the lowest component of ${\bf E}$  is the minus
 hermitian conjugate of   $\mybar E$:
\beq
{\bf E}= E + \theta \partial_{m} E \psi^{m}, \qquad 
\mybar E = - { E}^{\dagger} 
\eeq
Notice that the last term in the action is not integrated over the superspace, 
yet it is respectful to $\CQ=1$ supersymmetry.  It is not hard to show that 
 the $\theta$ component of the last term  is identically 
zero due to Jacobi identity, and translational invariance of the lattice 
action.  Hence it is supersymmetric. 

The dimensional reduction of the lattice action \Eq{lattice} 
to a single site by imposing 
the periodic boundary conditions (for example 
$\Lambda_{ \bf {n+  \hat e_{\mu}}}=\Lambda_{ \bf {n}}$) and by integrating out 
the auxiliary $d$ field  yields    the type IIB 
matrix theory action \Eq{typeIIB} 
expressed   in terms of $SU(5) \times U(1)$  multiplets   
 \beq 
S&&=\frac{\Tr}{g^2}\bigg[  ( \frac{1}{2}\sum_{m=1}^{5} \ [ \mybar z_m, z_m])^2 + 
\sum_{m, n=1}^{5}  |[z_m, z_n]|^2  
\cr \bigg. && \bigg. 
+ \lambda [\mybar z_m,  \psi^m ]  - \frac{1}{2}
{ \xi}_{mn}[ z^m,  \psi^n]  ) +  \frac{\sqrt 2 }{8} 
 \epsilon^{mnpqr}  \xi_{mn}
[\mybar z_p ,  \xi_{qr}] \bigg] \, .
\eqn{act2}
\eeq
This action is merely a rewriting of \Eq{typeIIB}, in terms of a  new set of
variables. Recall that  \Eq{typeIIB} is expressed in terms of $SO(10)$
multiplets, vector ${\bf 10}$ and   positive chirality spinor ${\bf 16}$. 
We could have obtained 
\Eq{act2}  from  \Eq{typeIIB}  by just employing  the decompositions  
 $ v_{\alpha}\sim {\bf 10} \rightarrow  z^m \oplus \bar z_m \sim  {\bf 5} \oplus 
{\bf \mybar 5}$  for bosons, and  
$ \omega \sim {\bf 16}   \rightarrow   \lambda \oplus \psi^m \oplus
\xi_{mn} \sim {\bf 1} \oplus {\bf 5} \oplus {\bf \mybar {10}}$ for fermions 
into $SU(5)$  multiplets as in  ref.\cite{Kaplan:2005ta}.  The 
dimensional reduction is used to set the notation and for later convenience. 

Notice that the bosonic matrix model takes a form similar to  
the potential in $d=4$ dimensional $\CN=1$ supersymmetric theories. There 
the potential is a sum of $d$ and $F$ (derivative of the superpotential) 
 term contributions  in the 
$\frac {d^2}{2} +  |F|^2$ form. In the zero dimensional matrix model 
expressed in manifest $\CQ=1$ superfields, the role of $F$ is  replaced with 
the $E$ functions and the bosonic action can be written as 
\beq 
S_{bosonic}=  \frac{d^2}{2} +  id  \, [ {\mybar z}_m, z^m ] + \sum_{m\neq n=1}^{5}|\mybar E_{mn}|^2 
\eeq
where $E^{mn}=[ \, z^m, z^n\, ]$.    
The deformation 
that we will construct in the next section will alter $E$, but not the 
$d$ term. 

\subsection{Dimensional reduction and twisted boundary conditions}

To formulate the deformed type IIB model, we start with the lattice 
action \Eq{lattice}
 of the $\CN=4$ SYM theory in four dimensional hypercube. Then 
we reduce  the lattice theory to a single site 
by imposing the twisted boundary conditions. 
This is indeed  the same philosophy used in  the construction of the 
twisted Eguchi-Kawai model \cite{Gonzalez-Arroyo:1982hz}. 
 
The difference between   deformed Type IIB and TEK is that  in $\CN=4$ lattice
action, there is a superfield associated with each $p$-cell. Hence, there are
many   gauge invariant plaquettes terms possible, which are  
shown in  fig.\ref{fig:flux}  and fig.\ref{fig:plaquette}, and will be 
explained 
momentarily.   
In TEK, we only have unitary link (1-cell) fields and the only possible 
gauge invariant plaquette terms are the square  plaquette terms. 

In reducing the lattice 
action \Eq{lattice} to a single point by using twisted boundary conditions,  
we request  periodicity modulo gauge rotations $g_{\mu}$. Let $\Psi$ be a   
a  generic field associated with  a $p$-cell on the lattice. We  utilize 
\beq
&&\Psi_{ \bf {n+  \hat e_{\nu}}}= g_{\nu} \Psi_{ \bf {n}} 
g_{\nu}^{\dagger} \, ,   \cr
&&\Psi_{ \bf {n+  \hat e_{\mu} + \hat e_{\nu}  }}=  
g_{\mu} g_{\nu} \Psi_{ \bf {n}} 
g_{\nu}^{\dagger} g_{\mu}^{\dagger} \, , \cr 
&&\Psi_{ \bf {n+  \hat e_{\mu} + \hat e_{\nu}  + \hat e_{\rho} }}=  
g_{\mu} g_{\nu} g_{\rho} \Psi_{ \bf {n}} g_{\rho}^{\dagger}
g_{\nu}^{\dagger} g_{\mu}^{\dagger} \, , \cr 
&&\Psi_{ \bf {n+  \hat e_{5}}}= g_5
 \Psi_{ \bf {n}}  g_{5}^{\dagger}, \qquad {\rm where} \qquad 
g_5 =  (\prod_{\nu=1}^{4} g_{\nu})^{\dagger}, 
\eqn{twisted}
\eeq
and then  set $\Psi_{ \bf {n}}$ to a site  independent value. 
The global  gauge rotation matrices $g_{\mu}$ satisfy  the 't Hooft
algebra \cite{'tHooft:1979uj} 
\beq
g_{\mu} g_{\nu} = e ^{i \Phi_{\mu \nu} } g_{\nu} g_{\mu}, \qquad 
{\rm equivalently},  \qquad  
e^{-i \Phi_{\mu \nu}/2 } g_{\mu} g_{\nu} = e ^{i \Phi_{\mu \nu}/2 }
 g_{\nu} g_{\mu}    \, .
\eqn{thooft}
\eeq
The antisymmetric $\Phi_{\mu \nu}$  matrix  is identified as the 
flux passing through the $\mu \nu$ plane. More precisely,  
$\Phi_{\mu \nu} \equiv  2 \pi \frac{n_{\mu \nu}}{N_c} $  where 
$n_{\mu \nu}$ is an integer associated with the integer   
 electric ($n_{4i}$) or magnetic ($n_{ij}$) flux modulo $\Nc$. 
 Also,  note that 
$\Psi_{ \bf {n-  \hat e_{\nu}}}= g_{\nu}^{\dagger} \Psi_{ \bf {n}} 
g_{\nu} $.  The second form of the equation  \Eq{thooft} is symmetric under 
the exchange
$\mu\leftrightarrow\nu$ and will be more useful in appendix. The commutation 
relation of $g_{\mu}$ with  $g_{5}$ is given by 
\beq
g_{\mu} g_{5} = e ^{i \Phi_{\mu 5} } g_{5} g_{\mu}, \qquad 
{\rm where }  \qquad   \Phi_{\mu 5}= -  \sum_{\alpha \neq \mu}^{4}
\Phi_{\mu \alpha}  
\label{eq:g5}
\eeq
The $\Phi_{\mu 5}$ is the  flux passing through  the 
plaquette whose 
boundary is an oriented parallelogram (generalization of the one
shown in fig.\ref{fig:flux} to four dimensions),  whose sides are 
the  link piercing the 4-cell, and the link along the  1-cell.
Such a plaquette has a 
nonvanishing projection 
on   three faces, and has zero projected area on the other three face which 
are 
orthogonal to the $\mu$  direction. Hence, the flux passing through it,  
is the
sum of fluxes passing through the faces, which this plaquette has
non-vanishing projection, i.e,  
$\Phi_{\mu 5}= -  \sum_{\alpha \neq \mu}^{4} \Phi_{\mu \alpha}$.  
 The two commutation relations, \Eq{thooft} and \Eq{g5}, can be combined 
into $ g_{m} g_{n} = e ^{i \Phi_{m n} } g_{n} g_{m} $ by keeping in mind  
the definition  of  $g_5$  and   $\Phi_{m n}$.

The reduction of the orbifold lattice action  \Eq{lattice} to a single site 
by using twisted boundary condition gives the action of the deformed Type IIB  
matrix model. The derivation of this  action is given in the appendix 
\ref{sec:ap1}. However, the deformed action is easy to understand on physical
grounds. Here we present a heuristic argument. 

First, note that the lattice action  \Eq{lattice} is a sum of a 
$Q$-exact and a  $Q$-closed expression.
 Schematically, both types are  
triangular plaquette terms whose
sides are bosonic (B) and fermionic (F) $\CQ=1$ supermultiplets.
 Upon reduction and appropriate field  redefinitions, 
all  the commutators will be replaced by ``deformed commutators'', where the 
deformation parameter is the flux passing through the corresponding 
triangular plaquettes:   
\beq 
&& S_{exact}  \sim { F [\;B_1, B_2 \; ]}|_\theta \rightarrow  
 F [e^{i ({\rm flux})} B_1 B_2 - e^{-i ({\rm flux })}
 B_1 B_2 ] |_{\theta}  \cr 
&& S_{closed} \sim  { F_1 [\;B, F_2 \; ]} \rightarrow  
 F_1 [e^{i ({\rm flux})} B F_2 - e^{-i ({\rm flux })} 
 F_2 B ] 
\eeq
Since there are $p$-form supermultiplets which reside on the
diagonal of each $p$-cell, there are a  number of possible plaquettes. 
The fluxes associated with each type of plaquette will be different.  
We first write down  the  deformed type IIB action in manifestly  
$\CQ=1$  superfields and  by preserving the  $U(1)^5$ global symmetry, 
 and then we explain the plaquette structures. The deformed action, which is the main result of this paper,   is given by 
\beq
&&S_{deformed} = \frac{1}{g^2} \Tr  \int  d \theta \bigg( -\frac{1}{2} 
{\bf \Lambda} 
{\partial}
_{\theta} {\bf \Lambda}  - {\bf \Lambda} ( [\mybar z_{\mu}, {\bf Z}^{\mu}] 
+ \frac{1}{24} \; [\mybar z^{\mu \nu \rho \sigma}, {\bf Z}_{\mu \nu \rho 
\sigma}]) \bigg. \cr &&
\bigg.  + 
\frac{1}{2}
{\bf \Xi}_{\mu \nu}( e^{-i \Phi_{\mu \nu}/2}  {\bf Z}^{\mu} {\bf Z}^{\nu} -
e^{i \Phi_{\mu \nu}/2}
{ \bf Z}^{\nu}{\bf Z}^{\mu})  \bigg. \cr
&& \bigg. 
+ \frac{1}{12}
{\bf \Xi}^{\nu \rho \sigma }( e^{i ( \Phi_{\mu \nu }+  
\Phi_{\mu \rho } + \Phi_{\mu \sigma }) /2}  
{\bf Z}^{\mu} {\bf Z}_{\mu \nu\rho \sigma} -
e^{-i ( \Phi_{\mu \nu }+  
\Phi_{\mu \rho } + \Phi_{\mu \sigma }) /2    }
{ \bf Z}_{\mu \nu\rho \sigma} {\bf Z}^{\mu})  \bigg) \cr
&&
+ \frac{\sqrt 2 }{2}    
 {\bf \Xi}_{\mu \nu} 
(e^{i (\Phi_{\rho \mu} +\Phi_{\rho \nu})/2}  \;  \mybar z_{\rho} \;  
  {\bf \Xi}^{ \mu \nu\rho }   
 - e^{-i (\Phi_{\rho \mu} +\Phi_{\rho \nu})/2} \;
 {\bf \Xi}^{ \mu \nu\rho }  
\mybar z_{\rho}) \cr
&&
 +
\frac{\sqrt 2}{8}   {\bf \Xi}_{\mu \nu}
(  e^{-i (\Phi_{\rho \mu} + \Phi_{\rho \nu} + 
\Phi_{\sigma \mu} +\Phi_{\sigma \nu} )/2} \;
\mybar z^{\mu \nu\rho \sigma} \, \,  {\bf \Xi}_{\rho \sigma} 
 -  
 e^{i (\Phi_{\rho \mu} + \Phi_{\rho \nu} + 
\Phi_{\sigma \mu} +\Phi_{\sigma \nu} )/2} \; 
{\bf \Xi}_{\rho \sigma} 
\mybar z^{ \mu \nu\rho \sigma}  )
 \qquad 
\eqn{deformed2}
\eeq
\\
In this expression, the penultimate and the last line are not separately 
supersymmetric, but their sum add up to a $Q$-closed form and supersymmetric.
The     $\CQ=1$ off-shell
supersymmetry transformations  can then be realized in terms of the following 
superfields, 
each of which is the reduction of the $p$-form field  to a single point 
\beq
{\bf \Lambda} &=& \lambda  -\theta   id , \cr
{\bf Z}^{\mu}  &=&  z^{\mu} + \sqrt 2  \theta \psi^{\mu}, \qquad \mybar
z_{\mu} , \cr
{\bf \Xi}_{\mu \nu} &=& \xi_{\mu \nu} -2 \theta  {\mybar E}_{\mu \nu}(\mybar z), \cr
{\bf \Xi}_{\mu 5} &=& \xi_{\mu 5} -2 \theta {\mybar  E}_{\mu 5}(\mybar z )  
\;\;  ({\bf \Xi}^{\nu \rho \sigma} ), \cr
{\bf Z}^{5}  &=&  z^{5} + \sqrt 2  \theta \psi^{5}  
 \;\; ( {\bf Z}_{ \mu \nu \rho \sigma} ), \qquad \mybar z_{5}   
 \;\; ({\mybar z}^{ \mu \nu \rho \sigma} ), 
\eeq
where  $\mybar z_m$ are supersymmetry singlets. 
The $\mybar E$ functions are given by 
\beq
{\mybar  E}_{\mu \nu}(\mybar z) &&= e^{-i \Phi_{\mu \nu}/2} \;   \mybar z_{\mu}
\mybar z_{\nu} - e^{+i \Phi_{\mu \nu}/2} \; \mybar z_{\nu}
\mybar z_{\mu} \cr
\cr
{\mybar   E}_{\mu 5}(\mybar z) &&= e^{i \sum_{\nu\neq \mu} 
\Phi_{\mu \nu}/2} \;  \mybar z_{\mu}
\mybar z_{5} - e^{-i \sum_{\nu \neq \mu }\Phi_{\mu \nu}/2} \; \mybar z_{5}
\mybar z_{\mu} 
\label{eq:Ebar}
\eeq
and they give rise to deformed  potential of the matrix theory. 
The $\mybar E$ functions are  supersymmetry singlets since they are functions
of $\mybar z$ and not $\bfz$, and are annihilated by the 
only supersymmetry in the theory: $Q \mybar E = 0$. 
Similarly, the  functions ${\bf E}$ are functions of general $\CQ=1$
  superfields, and  
are given by 
\beq
{\bf E}^{\mu \nu}= e^{-i \Phi_{\mu \nu}/2} \;   \bfz^{\mu}
\bfz^{\nu} - e^{i \Phi_{\mu \nu}/2} \; \bfz^{\nu}
\bfz^{\mu}   \cr 
{\bf E}^{\mu 5}= e^{-i \Phi_{\mu 5}/2} \;   \bfz^{\mu}
\bfz^{5} - e^{i \Phi_{\mu 5}/2} \; \bfz^{5}
\bfz^{\mu}  
\label{eq:Edef}
\eeq
Notice that, as before,  ${\bf E}$   has an expansion in superspace 
and  the lowest component of ${\bf E}$   is negative hermitian 
conjugate of   $\mybar E$, i.e; $\mybar E= - { E}^{\dagger} $.

The rest of this paper is devoted to the study of the deformed action 
\Eq{deformed2}.
Even though 
it is not apparent at first glance, it serves as a matrix regularization 
of the four dimensional $\CN=4$ SYM theory.   This identification requires 
neither   orbifold projection   
nor compactification  \cite{Taylor:1997dy}.  
We will show how  \Eq{deformed2} gives rise to the  four dimensional  
$\CN=4$ SYM theory, both commutative and non-commutative.  
In order to do that, it is important to understand the 
structure of the noncommutative moduli space, i.e,  identify the zero 
action configuration of the  bosonic potential and understand the
fluctuations around that background. Before moving on  we want to give a 
little bit more detail about the unusual plaquette structure and    
give a visual explanation for the deformations.

{\it Plaquettes}
 
The first line of \Eq{deformed2} 
 is the interaction of site superfield $\bfl$ with itself 
and with the link fields.  The links traverse 1-cell and the  
 4-cell diagonals 
in backward and forward directions and vice versa. 
As it does not surround any area,  the commutator
is unaltered. 

\EPSFIGURE[t]{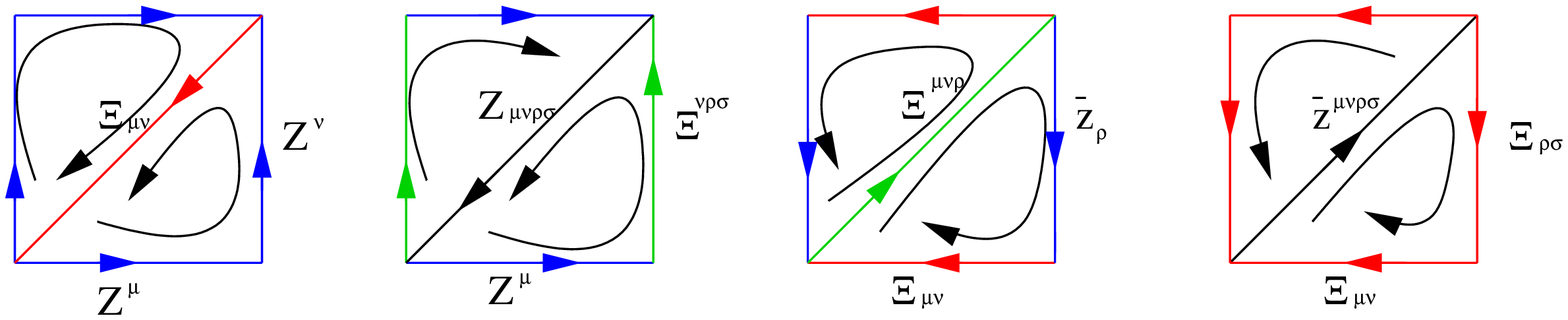,width=16cm}
{\sl  Plaquettes types are , respectively 2-1-1, 4-1-3, 3-2-1 
and 4-2-2 following the  name of the $p$-cell diagonal that the corresponding
superfield is residing.  The blue link is  link fields 
$\bfz_{\mu}, \mybar z_\mu$, red  is face diagonal (fermi) multiplet 
${\bf \Xi}_{\mu \nu}$, 
green is cube diagonal  ${\bf \Xi}^{ \nu \rho \sigma}$,  and black is   the
diagonal piercing the hypercube,  $\bfz_{ \mu \nu\rho \sigma},  
\mybar z^{ \mu \nu\rho \sigma}  $ 
\label{fig:plaquette}}

The second line of  \Eq{deformed2} is a signed sum over triangular 
face plaquettes.  
It is a face-link-link (2-cell, 1-cell, 1-cell) plaquette,  2-1-1 
for short.  The bosonic contribution of this term includes  the 
usual twisted Eguchi-Kawai (TEK). 
 This is most easily seen  by  using  the polar decomposition of the  
complex bosonic matrices and by looking to the angular fluctuations. 
As the triangular plaquettes  are oppositely oriented, 
the fluxes comes with opposite signs.   This is shown in 
fig.\ref{fig:plaquette}. 

The third line of \Eq{deformed2} is a signed sum of a gauge invariant product
of a cube-hypercube-link (or 3-4-1)
superfields. The 3-4-1 triangular plaquette has 
nonvanishing projection to only  
three faces of the hypercube, hence the deformation parameter is just the 
sum of these three fluxes. 

The penultimate line is a triangular plaquette of face-link-cube multiplets 
( or 2-1-3) and finally the last line is a face-hypercube-face (or 2-4-2). 
These are shown in the last two figures in fig.\ref{fig:plaquette}. The 
type 2-1-3 plaquette on a fixed Euclidean time-slice is shown in   
fig.\ref{fig:flux}.  The 2-1-3 plaquette in conjunction with 2-1-1 plaquettes 
provides further insides and is discussed below.

\subsection{ An intuitive explanation for the deformations}

\EPSFIGURE[t]{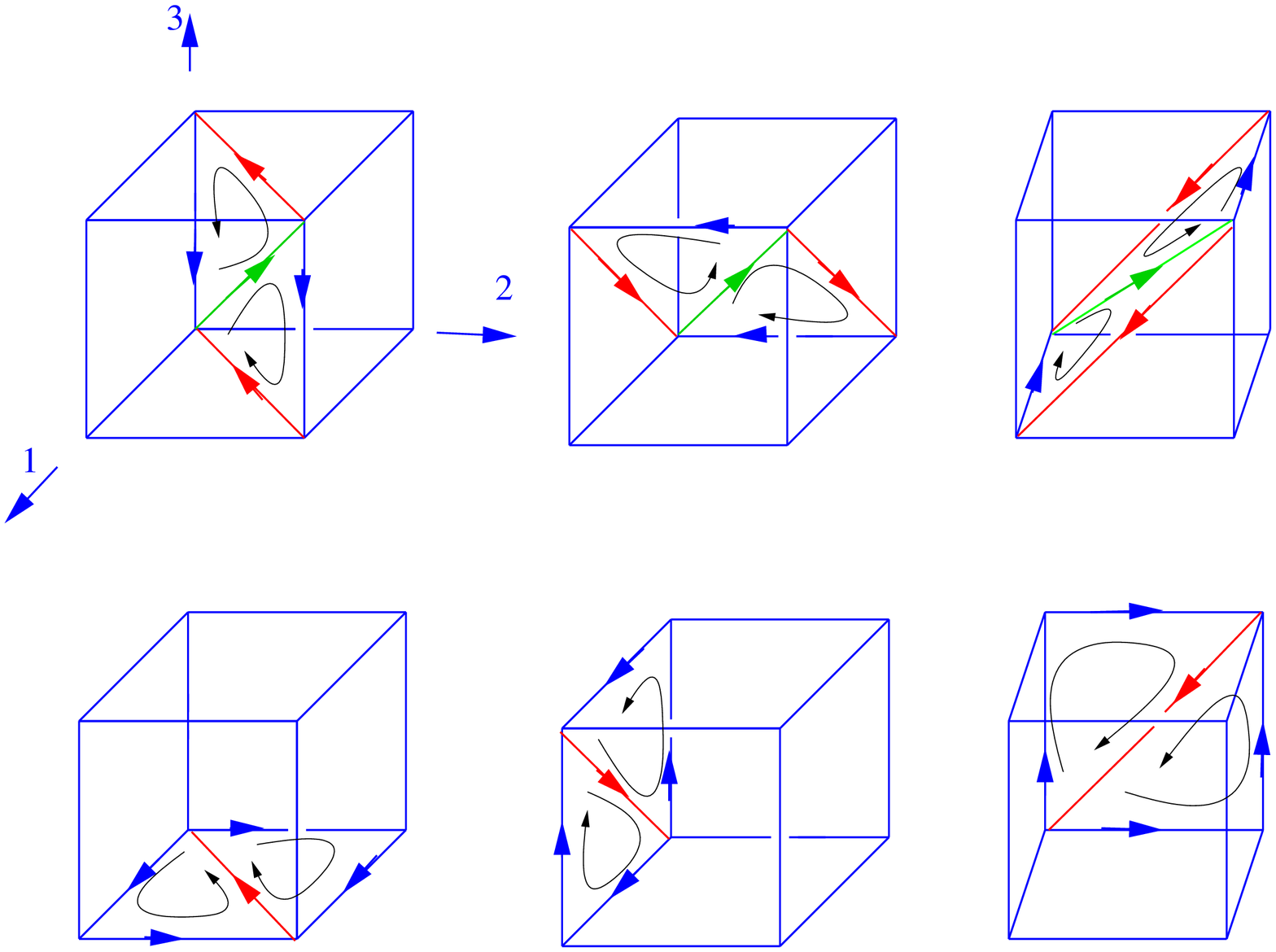,width=16cm}
{\sl 
The constant Euclidean time slice of the hypercubic lattice. All possible 
plaquettes
of type 2-3-1  and 2-1-1  are made explicit.  The flux through a 2-3-1 is the 
sum of fluxes passing through its projection on faces. The color coding is
same as fig.\ref{fig:flux}.    
\label{fig:flux}}%
Consider a fixed Euclidean time slice on the hypercubic lattice, which is  a
$d=3$ dimensional cubic lattice.  The background electric flux $\Phi_{i4}$ 
is parallel to the $x_4=$fixed hypersurface and does not generate 
a flux for the plaquettes living in that hypersurface. So the 
deformation of the action in  $x_4=$ fixed slice  
 will be purely due to magnetic flux $\Phi_{ij}$, where $i, j=1,2,3$.  
We will show how these 
fluxes arises by just looking at the orientation and position of the 
plaquettes in figure.\ref{fig:flux}.
 
Let us consider the part of the action that lives on the 
three dimensional hyperspace. The relevant terms in the action are 
${\bf \Xi}_{ij} {\bf E}^{ij}|_\theta +  
{\bf \Xi}^{kij}[\mybar z_k, {\bf \Xi}_{ij}]$ where 
${\bf E}^{ij}=[\bfz^i, \bfz^j]$. \footnote{Incidentally, 
this is also the  dimensional reduction  of the plaquettes  of our $d=3$ 
lattice action for $\CQ=8$ target theory. In three dimensional lattice, we
make the replacement 
 ${\bf \Xi}^{ijk}
\rightarrow \epsilon^{ijk} \chi$ and $\chi$ is a supersymmetry singlet.
See \cite{Cohen:2003qw}.}  
Both  are plaquette terms, but of different nature. The first one is the 
sum of the closed oriented loops on the three faces of the cube. 
The second one
involves the diagonal of the cube which pierces 
the  volume of the cube, a face diagonal, and an ordinary link. 
These two types are shown in the figure \ref{fig:flux}.  
The deformed action  is   
\beq 
{\bf \Xi}_{ij} [ e^{-i \Phi_{ij}/2} \bfz^i \bfz^j - e^{i \Phi_{ij}/2} \bfz^j 
\bfz^i ]|_\theta +  {\bf \Xi}^{ijk}[  e^{-i (\Phi_{ki} +\Phi_{kj})  /2}  
\mybar z_k {\bf \Xi}_{ij}  -   e^{i (\Phi_{ki} +\Phi_{kj})  /2} {\bf
  \Xi}_{ij} \mybar z_k  ]\; .
\eeq   

The orientation of the gauge invariant product of 
superfields (shown in figure \ref{fig:flux}) determine the sign of the
deformation parameter. The relative signs between 2-3-1 and 2-1-1 type 
plaquettes 
can also be understood similarly. For example, in figure \ref{fig:flux},   
any of the   2-3-1 
type plaquette has nonvanishing projection on two faces and zero projection 
on  the other. However, the orientation of the projected  2-3-1 is opposite
of the corresponding 2-1-1 plaquettes. This accounts for the overall
difference  of the sign among the fluxes between these two types of
plaquettes.  All the other plaquette terms and deformation parameters 
in deformed action \Eq{deformed2} can be understood similarly. 

 \section{Noncommutative moduli space }
\label{sec:3}
The classical moduli space of vacua is determined by the vanishing 
of the bosonic action. The bosonic action of the deformed theory, 
as expressed in  \Eq{Sb},  is a sum of the $d$-term and $E$-term. Therefore,
 the 
problem at hand reduce to the study of $d$  and  $E$-flatness, which is a set
of quadratic matrix polynomial equations \footnote{
The  full examination of the classical  moduli space of the arbitrary 
deformations of Type IIB
theory is beyond the scope of this paper.  
The deformations of the matrix model that we constructed  are of special type 
and  other deformations, which preserve more exact supersymmetry such as 
$\CQ=2, 4,8$ and bigger subgroups of the global $SO(10)$ R-symmetry,  
are possible as well. For example, the 
Fayet-Illiopoulis deformation,  mass 
deformation etc.
both of   which makes the moduli space noncommutative.
A classification and  analysis of
the deformations and their consequences   can probably
be performed along the lines of \cite{Berenstein:2000ux}
by applying the non-commutative algebraic geometry, and would be
interesting. 
There are also some  works exploring the deformations of Type IIB model 
\cite{Bonelli:2002mb} and its bosonic cousins \cite{Azuma:2005pm}.}. 

Surprisingly, the zero action configurations of the deformed Type IIB action 
turns out to be closely related to the TEK action, whose general solutions 
are well known  \cite{Das:1984nb, vanBaal:1985na}.  
This relation is somewhat  unexpected since  the the bosonic deformed 
Type IIB
action is expressed in terms of ten non-compact, algebra-valued  matrices,
whereas the TEK is expressed in terms of four group valued unitary matrices.  
The TEK, for a given nonzero deformation parameter, does not possess a 
moduli space, but a discrete set of vacua modulo gauge rotations.  This
discrete set of vacua play an essential role in the description of the 
 moduli space of deformed Type IIB. 
In the rest of this section  we first review the zero action configuration of
both EK and TEK, then explore the moduli space of the deformed Type IIB in
connection with TEK. 

\subsection{The relation  between TEK and deformed Type IIB } 
\label{sec:TEK}
The construction of the deformed Type IIB matrix model, given in appendix 
\ref{sec:ap1},  follows 
from a dimensional reduction scheme which admits  twisted boundary conditions. 
The reduction by using periodic boundary conditions  produces 
the Type IIB matrix model. 
This is in essence, the relation between  Eguchi-Kawai (EK) model and 
twisted Eguchi-Kawai (TEK) model. 

The EK model is a matrix model of $d$ unitary matrices $U_{\mu}$,
  obtained  by dimensionally reducing the lattice Wilson action to a 
single point. The action is typically written as  
\beq
S_{EK}= -\beta \sum_{\mu \neq \nu} \Tr (U_{\mu}U_{\nu} U_{\mu}^{\dagger}
U_{\nu}^{\dagger}-1 ) + h.c. =  
\beta \sum_{\mu \neq \nu} \Tr |U_{\mu}U_{\nu} -  U_{\nu}
U_{\mu} |^2  
\label{eq:EK}
\eeq
Historically, this model had been introduced to work the large $N_c$ limit 
of the gauge theories \cite{Eguchi:1982nm}. It is  shown in
\cite{Eguchi:1982nm} that the 
reduced model  
becomes equivalent to the full theory provided  the center symmetry (in the 
reduced model) is  not spontaneously broken. However, this does not happen to 
be the case  at weak coupling (large $\beta$) \cite{Bhanot:1982sh},
and there is indeed a remnant of the confinement-deconfinement 
phase transition at the level of the matrix model.
 At and above the critical value of the coupling constant $\beta_c$,
the Polyakov loop acquires a vacuum expectation value. The eigenvalues of the 
Polyakov loop  become nonuniformly distributed, and they clump.(For more 
recent  discussions of TEK in  context of large $N_c$ equivalences 
see \cite{Gonzalez-Arroyo:2005dz},  and in the noncommutative gauge theory 
context  see   \cite{Griguolo:2003kq}) 

The twisted EK is a variant of the EK model, which is introduced 
to prevent spontaneous symmetry breaking, hence is 
a cure for the clumping of eigenvalues. (Another solution is quenching
\cite{Bhanot:1982sh})
The action is altered in 
such a way that the eigenvalues of the unitary matrices are uniformly
distributed even at the very weak coupling. The TEK action is 
given by \cite{Gonzalez-Arroyo:1982hz} 
\beq
S_{TEK}= -\beta \sum_{\mu \neq \nu} \Tr (e^{-i\Phi_{\mu \nu}}U_{\mu}U_{\nu} 
U_{\mu}^{\dagger}
U_{\nu}^{\dagger}-1 ) + h.c. =  
\beta \sum_{\mu \neq \nu} \Tr |e^{-i\Phi_{\mu \nu}/2} 
U_{\mu}U_{\nu} - e^{i\Phi_{\mu \nu}/2} U_{\nu}
U_{\mu} |^2  \qquad
\label{eq:TEK}
\eeq
where  $e^{i\Phi_{\mu \nu}}$ is  the flux  factor as in \Eq{thooft}. 
Notice that the TEK action can be thought as a part of the full deformed 
Type IIB  action. This can be done by doing a polar decomposition of the 
complex matrices as $z_{\mu}= H_{\mu} U_{\mu}$ and looking to a configuration 
for which the radial modes are proportional to identity, as in 
\cite{Unsal:2005yh}. Freezing the radial fluctuations and paying attention to 
only angular ones yields 
$  \sum_{\mu \neq \nu}  |E^{\mu \nu}|^2 \sim S_{TEK}$, 
 where   $E$'s are given in  \Eq{Edef}.

The twisting (deformation) in \Eq{TEK} 
is an element of  ${\mathbb Z}_{N_c}$ and  has an important implication. 
It forces the eigenvalues 
of the Polyakov loops to be uniformly distributed even at weak coupling,
 preventing the Polyakov loops from acquiring vacuum expectation values. 
The zero action configurations of TEK action are given by 
\beq
U_{\mu}^0 \, U_{\nu}^0 = e^{i\Phi_{\mu \nu}} \, U_{\nu}^0 \, 
U_{\mu}^0
\eqn{minimaTEK}
\eeq
which has well-known twist-eating solutions. These solutions are expressed 
  in terms of  
clock $P_L$ and shift $Q_L$ matrices  given by 
$(P_L)_{jk}= e^{2 \pi i  k /L} \delta_{jk}$ and  
$(Q_L)_{jk}= \delta_{j+1, k}$. 

We consider the simplest deformations in $U(N_c)$ matrix  theory
corresponding to   $n$ 
units of electric and magnetic flux in z-direction. 
The flux matrix comes into the canonical form:
\beq
\Phi_{\mu \nu}=  \frac{2 \pi n}{N_c} 1_2 \otimes i \sigma_2 
\eeq 
where $1_2$ is two dimensional identity and 
$\sigma_2$ is the antisymmetric Pauli matrix.   
The zero action configuration for $N_c=L^2$ and $n=L$ background 
can be written as
\beq
U_{1}^0 = P_L \otimes 1_L, \qquad  U_{2}^0 = Q_L \otimes 1_L, \qquad 
U_{3}^0 = 1_L \otimes P_L, \qquad  U_{4}^0 = 1_L \otimes Q_L
\eqn{basis}
\eeq
The solution is  unique modulo gauge transformations. 
Its  significance  for the deformed type IIB
 will  be discussed in depth in the  next section, but at this point it is 
useful to realize that these are generators of a four dimensional (fuzzy) 
torus. 
 
\subsection{The noncommutative moduli space of deformed Type IIB }
\label{sec:solution}
The zero  action  configurations of the \Eq{deformed2} are  determined by 
the $d$ and $E$ flatness conditions  
\beq
&&id= \sum_{m=1}^{5}[\mybar z^{m}, z_{m}]= 0,  \qquad   \cr 
&&\mybar E_{\mu \nu}=  \mybar z_{\mu}
\mybar z_{\nu} - e^{i \Phi_{\mu \nu}} \; \mybar z_{\nu}
\mybar z_{\mu}=0  ,   \cr
&& \mybar E_{\mu 5}=    \mybar z_{\mu}
\mybar z_{5} - e^{-i \sum_{\nu \neq \mu }\Phi_{\mu \nu}} \; \mybar z_{5}
\mybar z_{\mu}=0  
\eqn{minima}
\eeq
which are a total of eleven  quadratic matrix polynomial equations. 
 The $d$-flatness  condition is unaltered with respect to undeformed case. 
The matrices satisfying the $E$-flatness condition will also satisfy 
the $d$-term condition, following the analysis of 
\cite{Luty:1995sd} . Henceforth, our  goal is to obtain the solution of 
the $E=0$. 
 
Let us first recall the moduli space of the (undeformed) Type IIB theory,
given by $d$ (as above) and  $E$-flatness conditions  
$E^{mn}= [z^m, z^n]=0 $.  
Solutions are mutually commuting, diagonal 
matrices with complex eigenvalues. Hence, the 
moduli space of the  matrix model \Eq{act2} with gauge group 
$U({\Nc})$  is 
\begin{equation}
\CM= {\mathbb C}^{5\Nc}/ {S_{\Nc}}
\end{equation}
where  ${S_{\Nc}}$ is the Weyl group of  $U({\Nc})$.
 This moduli space is a commutative one since the 
matrices are  simultaneously diagonalizable.  

We now  describe the noncommutative moduli space, given by  
\Eq{minima}.
The equation in the penultimate line of \Eq{minima} is, in 
fact, the same as the equation for the minima of the  twisted EK model 
\Eq{minimaTEK}, 
with one essential difference. 
In  TEK, the $U_{\mu}$  matrices are group valued, 
unitary matrices. Whereas, the $z_m$ matrices entering  \Eq{minima} are 
noncompact algebra-valued,   complex  matrices. 
The solutions of \Eq{minimaTEK}  will satisfy the \Eq{minima} as well. 
Then, the solution of  \Eq{minima}  will be of the form  
\beq 
z_{\mu} = c_{\mu} U_{\mu}^{0}, \qquad { \rm \, no \, sum}
\label{eq:mod1}
\eeq   
where $c_{\mu}$ is a complex moduli which is associated with the rigid
rotations and overall scaling of the eigenvalues. 
 The equation $E_{\mu5}=0$ will be satisfied if 
\beq
z_5= c_5 U_{5}^0 = c_5 \left( \prod_{\nu=1}^{4} U_{\nu}^0 \right)^{\dagger} 
\, ,
\label{eq:mod2}
\eeq
 where  $c_5$ is an independent  complex  number. The $z_5$, apart from the 
scale factor $c_5$, is not an independent matrix and its form is fixed by 
the other four $U_{\mu}^0$'s.      

\EPSFIGURE[t]{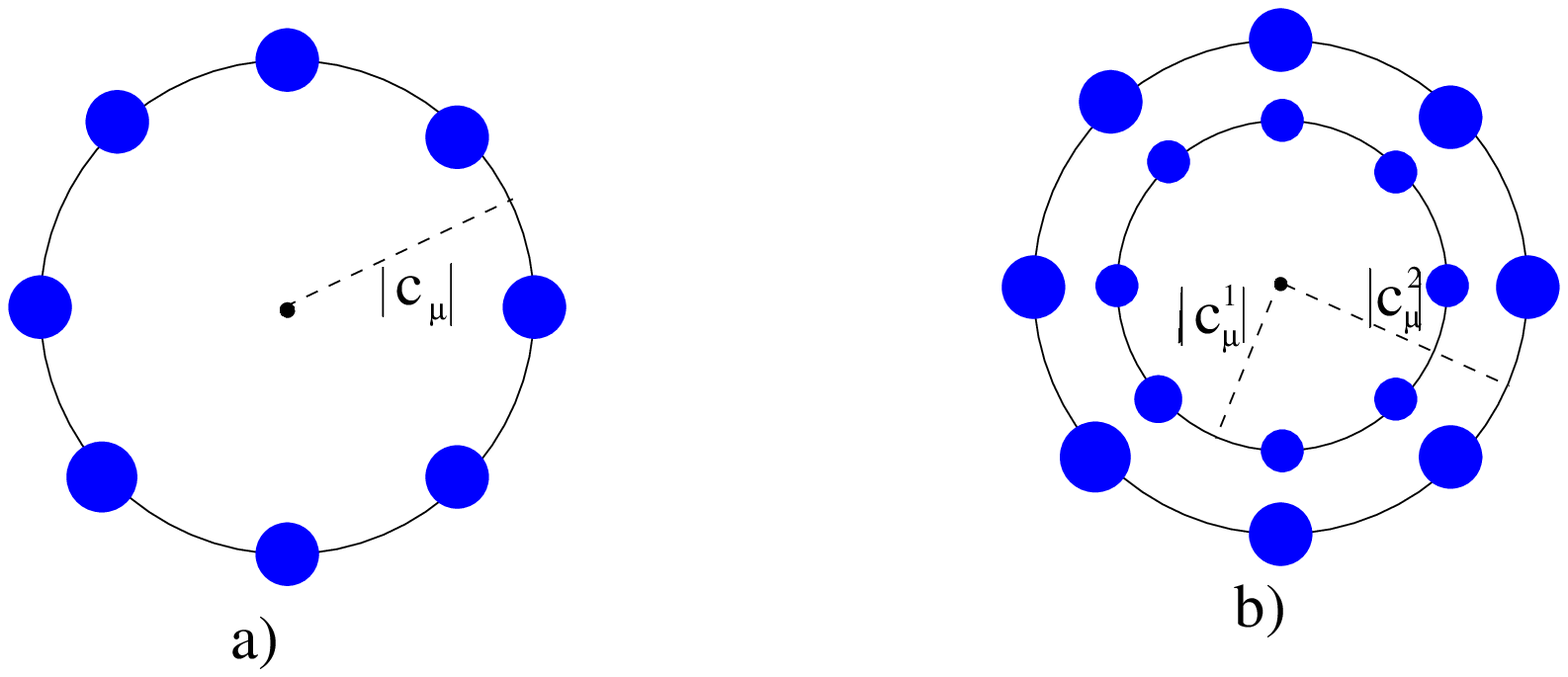, width=14cm}
{\sl The distribution of eigenvalues of  one of the  $z_{\mu}$ matrices.
  The moduli fields 
corresponds to the rigid rotations of eigenvalues and their overall 
scaling.  a) The $k=1$ case. The target theory  
has a  $U(1)$  gauge group.
b) The $k=2$ case.  For coinciding $c_{\mu}^1$ and $c_{\mu}^2$, there is an 
enhanced gauge symmetry group $U(2)$, otherwise the gauge group is 
 $U(1)\times U(1)$. 
\label{fig:moduli}}
The space of the minima of the deformed type IIB  action is  therefore 
parametrized by five continuous parameters, $c_1, \ldots ,c_5$   and 
a finite set of  unitary matrices satisfying the constraint 
\Eq{minimaTEK}.   The eigenvalues of the twist-eating matrices 
$U_{m}^{0}$ are roots of unity. 
 The eigenvalues of each $z_{m}$ are   uniformly distributed 
over a circle of radius $c_{\mu}$,  see fig.\ref{fig:moduli} a). The modulus
and the phase of 
 $c_m$ correspond to the overall scaling and the rigid rotation of the 
eigenvalues, as noted earlier.
To be more precise, for $N_c=L^2$, and flux $n=L$, the eigenvalues of the
matrix $z_m$ are
partitioned into $L$ groups, located at  $c_m e^{2 \pi ik/L}$ where $k=0,
\ldots L-1$.  
 For our interpretation of the deformed matrix model as a four
dimensional field theory, these moduli modes are essential 
\footnote{ The moduli space shown in figure \ref{fig:moduli} was raised  
 by Nima Arkani-Hamed almost a year ago in a discussion.  I thank him  
for sharing his ideas with me.}.    

Modulo these five complex numbers, 
and an extra unitary matrix $U_5^{0}$ (which is fixed
by the other four $U_{\mu}^{0}$),  
the  set of zero action configurations  of the deformed Type IIB  
matrix model \Eq{minima}  is 
same as the one of  the corresponding  TEK (with the same background flux). 
We can express the moduli space of the deformed theory as 
\beq
\CM_{deformed}= {\mathbb C}^5 \times M_{TEK} 
\eeq
Here  $M_{TEK}$ corresponds to a finite, discrete set of vacua  which 
happens to coincide  with the minima of corresponding TEK  \Eq{basis} 
As noted in section \ref{sec:TEK},  the procedure to construct  
$M_{TEK}$     is discussed in depth in literature and books  
(see for example \cite{Makeenko:2002uj}). 

The deformation 
changes the structure of the classical moduli space  
drastically \footnote{ 
For special values of flux, such as $\Phi_{\mu\nu}= 0$ except for   
$\Phi_{12}$,
the moduli space becomes (locally)  
${\mathbb C}^{2\Nc} \times  {\mathbb C}^{3} \times  M_{TEK2}$ 
i.e, the eigenvalues in the $z_3$ and $z_4$ start to move freely again
and $M_{TEK2}$ is the zero action configuration of two dimensional TEK.   
This deformation can be shown to be relevant to construct the 
$\CN=(8,8)$ theory in two dimensions.
Finally, when all the deformations are turned off, we recover  
$\CM= {\mathbb C}^{5\Nc}/ { S_{\Nc}}$
},  turning the  classical commutative moduli space  
 $\CM= {\mathbb C}^{5\Nc} /  S_{\Nc}$
in which 
the eigenvalues can
take any value (without increasing the action) 
to a  space in which  there are only five free complex
parameters. 
Recalling the polar decomposition of a complex $GL(N_c, {\mathbb C})$ 
into  a product of a Hermitian positive 
definite matrix  and a unitary matrix, $z_{\mu}=H_{\mu}U_{\mu}$,   
we observe that,  in the moduli space of the  deformed theory, the radial 
coordinates are forced to be equal: $H_{\mu}= |c_{\mu}| 1_{N_c}$.
The nonuniformities  disrespecting the $\mathbb Z_L$ in the eigenvalue 
distribution, in  either  radial  or angular direction, increase the
action.   

The reader may also wonder if   there are  other solutions to 
\Eq{minima} and if any, 
what do they correspond to.  There are indeed such solutions and  the ones  
 for which at least 
two out of five $z_m$ are zero are of different character.  
For example, in  $z^1=z^2=0$ case,   
it is not possible to generate a four dimensional  spacetime.  As we will see
in section \ref{sec:fluc}, the main moral of \Eq{mod1} is that it forms a 
complete basis for $GL(N_c, {\mathbb C})$ matrices, which in turn becomes the 
basis for the emergent lattice theory. The case for which at least two out of
five matrices is zero a full basis for $GL(N_c, {\mathbb C})$ can not be
constructed and we do not obtain a four dimensional target theory. However,
it is still possible to obtain a two dimensional target theory. If at least 
four out of five matrices are zero, then  it is not even possible to obtain a 
two dimensional theory.  The technique for the study of the full set of 
solutions is given in  ref.\cite{Berenstein:2000ux},  and a similar
deformation of the superpotential of $\CN=4$ is worked out there as an
example. We refer the reader to ref.\cite{Berenstein:2000ux} for further 
discussions.

{\it (Ir)reducible Representations}

Up to this point, we have discussed the theory with  $N_c=L^2$ and $n=L$, 
where $\Phi_{\mu \nu}=  \frac{2 \pi }{L} 1_2 \otimes i \sigma_2$.  As we will
see momentarily, this background 
supports  a $U(1)$ target 
theory on  a four dimensional lattice with $L^4$ sites.  
In order to get  a $U(k)$ theory on the same lattice, all we need to do is 
to take  $N_c=L^2k$ and $n=Lk$.  In this case, the moduli space is described 
by  matrices 
\beq 
z^{m} = C_{m}  \otimes  U_{m}^{0}, \qquad   {\rm where} \qquad 
 C_{m}= {\rm Diag} \; ( c_m^{1}, \ldots,  c_m^{k})  
\label{eq:modk}
\eeq 
where $C_m$ is diagonal $k \times k$  moduli matrix 
and the solution  can be thought as a direct sum of irreducible 
solutions in the form $ \bigoplus_{i=1}^k  c_m^i U_m = C_{m}  \otimes
U_{m}^{0}$.
Therefore,  the moduli space can be expressed as   
\beq
\CM_{deformed}= {\mathbb C}^{5k}/{S_k} \otimes M_{TEK} 
\eeq

The four dimensional lattice basis in terms of zero 
action configurations is encoded into  $M_{TEK}$, and 
   will be given explicitly in the next section.  We will see how the 
noncommutativity of  the moduli space is tied to the noncommutativity 
of lattice base space.   

Before moving on, we want to make one more  observation.  
 Geometrically,  the $k=1$ describes the eigenvalues residing on the 
surface of a four dimensional (fuzzy) hypertorus $T^4$. 
For $k \geq 2$, there are multiple, concentric fuzzy tori $T^4$, 
see fig.\ref{fig:moduli}b).  
When these concentric tori are on top of each other, there is an enhanced 
gauge symmetry group $U(k)$. If all of them are separated, the gauge group 
of the lattice theory is $U(1)^k$. This phenomena is same as 
 the emergence of $U(k)$ gauge theory on a stack of $k$ D-branes wrapped on 
a hypertorus.  We will   comment more on that in section \ref{sec:5}.

\section{The emergence of (noncommutative) base space}
\label{sec:4}
\subsection{Fluctuations}
\label{sec:fluc}
In this section, we  show  that the deformed  Type IIB action 
\Eq{deformed2}, without any orbifold constraints whatsoever,  is a lattice 
regularization of $\CN=4$  SYM theory.  The emergent lattice 
is noncommutative. However, depending on the deformation parameter, 
it is possible to find two continuum limits: one  is regular $\CN=4$ SYM, and
the other is its noncommutative counterpart. 

Let us start with the  deformed 
Type IIB  action \Eq{deformed2} with gauge group $U(N_c)$  where 
$N_c=L^2k$ and  with the  flux matrix $\Phi_{\mu \nu} = \frac{2\pi}{L} 
(1_2 \otimes i\sigma_2)$. In  section \ref{sec:solution}, we found 
the zero action 
configuration of deformed type IIB. We will analyze the bosonic and fermionic
fluctuations around this background. To do that, we decompose the bosonic 
matrices into  fluctuations  and  background  as  
\beq 
z^{m}= {\widetilde z_{m}} U_{m}^{0}  
\equiv    {\widetilde z}^{m} D_{m}  \qquad {\rm no \;\; sum} 
\eqn{fluctuations}
\eeq 
where $U_{m}^{0}$ is a background field configuration and 
$ {\widetilde z^{m}}$ are fluctuations around it.  The change in naming of the
background  is due to the physical interpretation, that the $U_{m}^{0}$ will
have momentarily, as displacement (discrete translation) operator on the 
lattice. 
Substituting \Eq{fluctuations} into the bosonic part of the deformed action
\Eq{deformed2}  yields
\beq 
 S_b=\frac{\Tr}{g^2} \left[ \frac{1}{2} \Big( \sum_{m=1}^{5} 
   [(D_m^{\dagger} \widetilde {\mybar z}_m D_m) 
( D_m^{\dagger} \widetilde z^m D_m)  -  \widetilde z^m 
\widetilde{\mybar z}_m] \Big)^2    + 
\sum_{m,n}|\widetilde z^m (D_m{\widetilde z}^n D_m^{\dagger}) -
m \leftrightarrow n  |^2  \right] \;\; \;\;
\eeq  
where the phase factors disappeared upon the use of the  \Eq{minimaTEK}.
 
The fluctuations  entering the action are  $U(N_c)$ algebra valued and are 
conventionally  expressed  as $\widetilde z=  z^A t^A ; 
\,\, A= 1, \ldots N_c^2 $; where  
$t^A$ are Hermitian 
generators and satisfy the standard commutation relations 
$[t^A, t^B]= i f^{AB}_{ {\phantom{AB} }  C} \,  t^C$ .   Instead, we will 
use a basis 
which makes the mapping of the $N_c\times N_c$ (where $N_c=L^2k$)  
 matrix  into $U(k)$ fields 
on $L^4$  lattice clearer. We introduce   $J_{\bf p}=  
e^{-i \pi (p_1 p_2 + p_3 p_4)/L}   (U_2^0)^{p_1}  
(U_1^0)^{p_2}  (U_4^0)^{p_3}  (U_3^0)^{p_4}  $,  
where $p_{\mu}=1, \ldots L$  which  form a complete and 
orthogonal set  for $L^2 \times L^2$ matrices, $GL(L^2, \mathbb C)$.   
The $J_{\bf p}$ basis 
is identifiable as the dual momentum basis for the lattice and {\bf p} is 
an integer vector  labeling points in Brillouin zone, $\mathbb Z_L^4$. 
By a Fourier transform, we construct the coordinate basis  
$\Delta_{\bfn}$, where 
\beq 
\Delta_{\bfn}= \frac{1}{L^4} \sum_{ {\bf p} \in {\mathbb Z}_L^4} J_{\bf p} e^{2 \pi i{\bf p}.{\bf n}/L}
\eqn{realspace}
\eeq
 Hence, 
we can express the map between a $\widetilde z$ fluctuation 
matrix and fields in the $L^4$ dual 
momentum space lattice  (or equivalently $L^4$ real space lattice ) as 
\beq
\widetilde z=\sum_{A=1}^{N_c^2} z^A t^A = 
\sum_{{\bf p} \in {\mathbb Z}_L^4} \sum_{a=1}^{k^2} 
 ( t^a \otimes J_{\bf p}^{\dagger}) z_{\bf p}^a  =  \sum_{{\bf p} \in
   {\mathbb Z}_L^4} 
z_{\bf p} \otimes J_{\bf p}^{\dagger}
\qquad {\rm or } \qquad  
{\widetilde z}= \sum_{{\bf n} \in {\mathbb Z}_L^4} z_{\bf n} \otimes  \Delta_{\bf n} 
\eqn{mat-field}
\eeq
We are essentially  expressing a $GL(L^2k, {\mathbb C})$ matrix by using 
complementary complete basis matrices, $J_{\bf p} \in GL(L^2, {\mathbb C})$
and $t^a \in GL(k, {\mathbb C})$. 
Notice that this representation makes 
clear how color degrees of freedom of the deformed matrix theory turns into 
 a noncommutative space time in the lattice theory.
   The properties of the   basis matrices are given  in 
\cite{Makeenko:2002uj}.   
The beauty of this basis is that the zero-action configurations 
 $D_{\mu} \equiv  U_{\mu}^0$ acts 
as displacement  operator on the position space lattice. Namely, 
\beq
D_{\mu} \left(\Delta_{\bf n} \otimes 
z_{\bf n} \right)D_{\mu}^{\dagger} = \Delta_{\bf n- \hat e_{\mu} } \otimes z_{\bf n} 
\eqn{shift}
\eeq
This relation  provides  a nearest neighbor interaction  
on the lattice. By using the orthogonality of basis matrices   
$J_{\bf p}$  and \Eq{shift}, the  bosonic action can be expressed as  
\begin{eqnarray}
S = \frac{\Tr}{g^2} \sum_{{\bf n} \in {\mathbb Z}_L^4 } \left[ \frac{1}{2} 
 (\sum_m (\mybar z_{m,\bf {n - \hat e_m}} 
 \star { z}_{m,\bf {n - \hat e_m}}
 - { z}_{m, \bf n}\star  \mybar z_{m, \bf n}))^2
+  
\sum_{m,n} |({z}_{m, \bf n} \star  
{ z}_{n, \bf {n + \hat e_{m}}}- m \leftrightarrow n
|^2 \right] \qquad 
\eqn{latticeb}
\end{eqnarray}
This is indeed just the noncommutative generalization of bosonic part of 
 \Eq{lattice}.   
Performing the same analysis at the level of superfields, it is easy to 
show that the deformed Type IIB action  \Eq{deformed2} may be rewritten as 
\begin{eqnarray}
S &=& \frac{\Tr}{g^2} \sum_{\bf n}  \int \;  d \theta \bigg( 
-\frac{1}{2} {\bf \Lambda_{\bf n}}\star  {\partial}_{\theta} 
{\bf \Lambda_{\bf n}}  
- {\bf\Lambda_{\bf n}}\star (\mybar z_{m,\bf {n - \hat e_m}} 
 \star {\bf Z}_{m,\bf {n - \hat e_m}}
 - {\bf Z}_{m, \bf n}\star  \mybar z_{m, \bf n} ) \bigg. \cr
&&  \bigg. 
+ \frac{1}{2} 
{\bf \Xi}_{\mu \nu, \bf n} \star ({\bf Z}_{\mu,\bf n} \star  
{\bf Z}_{\nu, \bf {n + \hat e_{\mu}}}- 
{\bf Z}_{\nu, \bf n} \star
 {\bf Z}_{\mu, \bf {n + \hat e_\nu}})+ 
{\bf \Xi}_{\mu 5, \bf n} \star ({\bf Z}_{\mu,\bf n} \star 
{\bf Z}_{5, \bf {n + \hat e_{\mu}}}- 
{\bf Z}_{5, \bf n} \star
 {\bf Z}_{\mu, \bf {n + \hat e_5}}) 
\bigg) \cr
&&+  \frac{\sqrt 2 }{2}  \epsilon^{\mu \nu \rho \sigma
} {\bf \Xi}_{\mu \nu , \bf n} \star
(\mybar z_{\rho, \bf {n-  \hat e_{\rho}}} \star 
{\bf \Xi}_{\sigma5, \bf {n +  \hat e_{\mu} + 
\hat e_{\nu} }} - 
{\bf \Xi}_{\sigma 5, \bf {n -  \hat e_{\sigma} - \hat e_5 }} 
\star \mybar z_{\rho, \bf {n +  
\hat e_{\mu} + \hat e_{\nu} }} ) \cr
&&+
\frac{\sqrt 2 }{8}  \epsilon^{\mu \nu  \rho \sigma} {\bf \Xi}_{\mu \nu, 
\bf n} \star
(\mybar z_{5, \bf {n-  \hat e_5}} \star   {\bf \Xi}_{\rho \sigma , 
\bf {n +  \hat e_{\mu} + 
\hat e_{\nu} }} - 
{\bf \Xi}_{\rho \sigma, \bf {n -  \hat e_{\rho} - \hat e_{\sigma }} }
\star \mybar z_{5, \bf {n +  
\hat e_{\mu} + \hat e_{\nu }}})
\eqn{lattice2}
\end{eqnarray}
The lattice star product is defined by 
\beq
\Psi_{1,\bfn} \star  \Psi_{2,\bfn} =  \frac{1}{L^4}\sum_{{\bf j}, {\bf k}}
\Psi_{1,{\bf j}} \; \Psi_{2,{\bf k} } \;\; e^{-\frac{4 \pi i} {L^2 \theta'}
  \;  ({\bf j}- \bfn) \wedge   
({\bf k}- \bfn) } 
\eeq
which is  a non-local product and  $\theta'= 2/ L$ is a
dimensionless noncommutativity parameter on the lattice. 

\subsection{Classical spectrum}
The action  \Eq{lattice2} has no explicit lattice 
spacing, but it has a classical moduli space for which the bosonic 
lattice action
vanishes.  The distance from the origin of the moduli space is identified as 
inverse lattice spacing, and hence the continuum limit is defined as a 
trajectory out to infinity in the moduli space. 

The zero action configuration of the lattice 
 corresponds to site independent diagonal matrices  
 $z_{m, \bfn } = C_{m}= {\rm Diag} \; ( c_m^{1}, \ldots,  c_m^{k})  $, 
hence the moduli space is   given by $\CM =  \mathbb C^{5k}/S_k$.  Notice
that  this is 
(and has to be)  the moduli space of the  deformed 
Type IIB action, as  \Eq{lattice2} is merely a rewriting. 

We expand the action around  a $U(k)$ gauge symmetry  preserving, 
 hypercubic spacetime lattice point:
\beq 
 z^{\mu}_{\bfn } = \mybar z_{\mu, \bfn} =  
\frac{1}{\sqrt2 a} 1_k,  \qquad  z^{5}_{\bfn } = \mybar z_{5, \bfn } = 0
\label{eq:point}
\eeq
We use a cartesian  decomposition  of  the  bosonic
 matrices around the point \Eq{point} as
\beq
z_{\mu, \bfn}= \frac{1}{\sqrt 2 a} 1_k +  
\frac{S_{\mu, \bfn} + i V_{\mu, \bfn}}{\sqrt 2}, \qquad   
z_{5, \bfn }=  \frac{S_{5, \bfn } + i S_{6, \bfn }}{\sqrt 2} 
\eeq 
where $S$ and $V$  are identified as six scalars and four gauge 
 bosons of the target   $\CN=4$ supersymmetric Yang-Mills theory.

The classical spectrum can be found by  diagonalizing 
the kinetic terms, similar to the appendix B in \cite{Kaplan:2005ta}. 
This can be done by introducing the Fourier transform of 
a field $\Psi_{\bfn}$
\beq 
\Psi_{\bfn} = \frac{1}{L^2} \sum_{{\bf p} \in B.Z.}
e^{i {\bf p.n}a}\Psi_{{\bf p}}  
\eeq
where the $ p_{\mu}= \frac{2 \pi}{La} {\hat p}_{\mu} \in [-\pi/a, \pi/a) $
is discrete momenta in  hypercubic Brillouin 
zone (B.Z.), and ${\hat p}_{\mu}$ is integer in the range $[-L/2, L/2)$.  
 This diagonalize 
the kinetic terms for the bosonic action
\beq 
\frac{\Tr}{2g^2} \sum_{{\bf p} \in B.Z.}  \big[ \sum_{a=1}^{6}S_{a, {\bf p}} M_{\bf p}^2   
S_{a, {- \bf p}} +   V_{\mu, {\bf p}} 
G_{\mu \nu, {\bf p}}   
 V_{\nu,- {\bf p}} \big]
\eeq
where we define
\beq
{\CP}_{\mu} \equiv  \frac {2}{a} \sin \; \frac{a p_{\mu}}{2}\, ,  \qquad 
M_{\bf p}^2 = \sum_{\mu=1}^{4}  {\CP}_{\mu}^2 \, ,   \; \qquad\; 
 G_{\mu \nu, {\bf p}} = M_{\bf p}^2 \delta_{\mu \nu} -   {\CP}_{\mu}
    {\CP}_{\nu} e^{-i a (p_{\mu} - p_{\nu})/2} \qquad 
\label{eq:spectrum}
\eeq
The six scalars are degenerate for a given momenta and have eigenvalue  
$M_{\bf p}^2$. 
The eigenvalues for the  gauge boson matrix   $G_{\mu \nu, {\bf p}}$ 
are  a three fold degenerate $M_{\bf p}^2$  and zero. The zero mode of $G$ is 
the
consequence of  gauge invariance.  The essential point here is the absence 
of doublers in the bosonic spectrum,  $M_{\bf p}^2$,  which is only zero at the
origin of the Brillouin zone and which reaches  its maximum at the corners.
The exact supersymmetry implies there are no fermion doublers either 
(see   \cite{Cohen:2003aa} for an explicit analysis.) 
The absence of the fermionic doublers is also a consequence of associating 
the sixteen fermions with the $p$-cells of the lattice 
(see \cite{Rabin:1981qj}).
\footnote{Notice that the spectrum \Eq{spectrum} is not the same as 
the spectrum of $A_4^*$ lattice, given in \cite{Kaplan:2005ta}.  This is due to
difference in the  lattice structures, ( $A_4^*$  versus hypercubic ), and
consequently their Brillouin zones ($A_4$ versus hypercubic).  
This is analogous to the 
difference in the acoustic phonon spectrum
of body-centered cubic and simple
cubic lattices in solid state physics.
 However, in our case, the long-wavelength limit of  these two lattice theories
coincides and become the  $\CN=4$ in four dimensions.}.
In the continuum limit, 
which is defined by 
\beq 
a\rightarrow 0,\;\;\;  L\rightarrow \infty, \;\;\; La= {\rm fixed},
\eeq
we obtain   
${\CP}_{\mu} \rightarrow p_{\mu}$ and $M_{\bf
  p}^2\rightarrow p^2 $ and $G_{\mu \nu, {\bf p}} \rightarrow 
(p^2 \delta_{\mu \nu} - p_{\mu} p_{\nu})$  correctly reproducing the kinetic
terms of bosonic action. 

{\it Moduli space: A second pass}
 
We reconsider the eigenvalue distribution shown in 
fig.\ref{fig:moduli} and make various observations. 
For simplicity, consider the case $k=1$.  In finding the spectrum, we expand 
the fields around the point \Eq{point} in moduli space. This means  
setting  $c_\mu =  \frac{1}{a}$ and $c_5=0$, where $a$ is interpreted  
 as the lattice spacing.  
Turning on $c_5$  is related to deforming the hypercubic lattice to the
$A_4^*$  lattice and tilting the $T^4$, and we  ignore this possibility 
 for  simplicity. 
Consider the eigenvalue distribution on the circle shown in 
fig.\ref{fig:moduli}, which is one of the
directions on fuzzy $T^4$ on which eigenvalues are distributed.
The  circumference of the circle is 
\beq 
2 \pi c_{\mu} = \frac{2 \pi}{a} 
\eeq
which is, in fact, the extent of the Brillouin zone in the $\mu$ direction. 
The  separation between two nearest-neighbor in moduli space
 along the circle   (geodesic  distance on the torus) is given 
by $2 \pi c_{\mu}/L = \frac{2 \pi}{La}$  
and this corresponds to the   minimal lattice momenta. The geodesic 
separation between 
two eigenvalues separated by $\hat p_{\mu} \in [-L/2, L/2)$ units is given by  
$p_{\mu}= \frac{2 \pi}{Na}  \hat p_{\mu} \in [- \pi/a, \pi/a)$ and  is the
discrete momenta in Brillouin zone of hypercubic lattice. As a last remark, 
notice that the external distance between the same  two points as above is  
$\frac{2}{a}
\sin \frac{a p_\mu}{2}$, the acoustic phonon spectrum in one dimension. 
The continuum limit is taken as a trajectory out to infinity in moduli space, 
by keeping the volume and four dimensional coupling constant fixed. Hence, 
the circumference  of eigenvalue circle in moduli space (and extent 
of the Brillouin zone) becomes
$p_{\mu} \in [-\pi/a, \pi/a) \rightarrow [-\infty, 
\infty)$.  It seems natural to interpret the four dimensional torus that the
eigenvalues reside as the Brillouin zone, and the action 
 as the momentum space action, with momenta summed over the Brillouin
zone.  

In the derivation of \Eq{lattice2}, we used a Fourier transform, which took
us from momentum space representation ($J_{\bf p}$) 
of the lattice to the position space representation  ($\Delta_{\bfn}$).
Hence, the inverse extend of Brillouin zone (divided by $2\pi$ ) becomes 
the UV cutoff. The  inverse of minimal momenta (divided by $2\pi$) 
becomes the IR cutoff ($La$), the size  of the spacetime torus.   
The four dimensional target  field theory lives on this spacetime torus,
whose extent remains fixed as we take the continuum limit.  It should be 
emphasized that the target space is not the space that the eigenvalues reside, 
but its Fourier transform. 

\subsection{Localization of interactions} 
In this subsection, we will describe  the circumstances under which the
non-locality in  \Eq{lattice2} disappears and the interaction becomes 
local in spacetime.  Upon that, the continuum limit of the action becomes the 
same as the continuum limit of \Eq{lattice} and reproduces the $\CN=4$ 
SYM in four dimensions. Since the calculations  are very similar 
to   \cite{Kaplan:2005ta}, they will  not be duplicated here. At the end, 
we also comment briefly on the non-commutative spacetime limit. 

The interaction terms of  the lattice action  \Eq{lattice2} 
are non-local, involving a sum over all lattice points.  
In the continuum limit,  the star-product becomes 
\beq
\Psi_1 ({\bf w})\star \Psi_2 ({\bf w}) = \int d^4 x \; d^4 y  \;
\Psi_1 ({\bf x})  K( {\bf x} - {\bf w}, {\bf y} - {\bf w} )   
\Psi_2 ({\bf y})
\label{eq:contstar} \; .
\eeq
The kernel of the integral  is   
\beq
K( {\bf x} - {\bf w}, {\bf y} - {\bf w} ) = \frac{1}{\pi^2 \det \Theta} 
e^{-2i ({\bf x} - {\bf w})_{\mu}  (\Theta^{-1})^{\mu \nu}
 ({\bf y} - {\bf w})_{\nu}}
\eeq 
and   the dimensionful noncommutativity parameter  is given   by
\beq
\Theta_{\mu \nu}= \frac{L^2 a^2 \theta'}{2 \pi} 
(1_2 \times i \sigma_2)_{\mu \nu} .  
\eeq

Let us consider the continuum limits  in which the  volume of the target space 
is kept fixed.  The length scale associated with noncommutativity  is the 
square root of $\Theta_{12}$, 
denoted as $\sqrt {\Theta}$ for short.   
Then, for $\sqrt {\Theta}/ (La) \rightarrow 0  $, i.e, the limit at which
noncommutativity goes to zero,   the kernel localizes sharply and becomes a
Dirac delta function:
\beq
K( {\bf x} - {\bf w}, {\bf y} - {\bf w} ) \rightarrow \delta
( {\bf x} - {\bf w}) \delta({\bf y} - {\bf w}),   
\eeq
 yielding a
local product out of  the star product. From the spacetime point of view,
trading  the nonlocality of the product of functions back to spacetime
noncommutativity, this limit yields a commutative spacetime.  
This is  certainly an interesting limit as 
it reproduces  the ordinary  $\CN=4$  SYM  target   theory.

Explicit calculation shows that, in the continuum limit, we obtain the action  
of $\CN=4$ SYM  at tree level.
\beq
S =\frac{1}{g_4^2} \,\int d^4x  \Tr
\Bigl(\frac{1}{4} V_{\mu\nu} V_{\mu\nu} + \frac{1}{2} (D_\mu S_a)^2
-\frac{1}{4}\left[S_a,S_b\right]^2 
+\frac{1}{2}  \omega^T\,C \, ( D_\mu \, \tilde \Gamma_\mu 
\,  \omega + 
i  \tilde\Gamma_{4+a}\, [S_a,
\, \omega])  \Bigr) \qquad 
\eqn{target4}\eeq
where $D_{\mu}= \partial_{\mu} + i [ V_{\mu},  \cdot \; ]$ is the 
covariant derivative and $V_{\mu \nu}= -i [D_{\mu},
D_{\nu}]$ is the nonabelian field strength. 
The $SO(10)$ gamma matrices ${\tilde  \Gamma}_{\alpha}$ are basically a
  reshuffling of the  gamma matrix basis introduced in \Eq{typeIIB}. 
They are given as  $ {\tilde  \Gamma}_{\mu}=  \Gamma_{\mu}, \; 
{\tilde  \Gamma}_{4+m}=  \Gamma_{5+m}, \; {\tilde  \Gamma}_{10}=  \Gamma_{5}$,
following the conventions of section 2.2   of \cite{Kaplan:2005ta}.
The  charge conjugation matrix $C$ is unchanged with respect to 
\Eq{typeIIB}.  

 Even though the deformed type IIB theory
\Eq{deformed2} , hence
lattice theory \Eq{lattice2}, possesses 
only  $\CQ=1$ exact supersymmetry  and  a $U(1)^5$  global symmetry,  
the target theory has  full  $\CQ=16$  supersymmetries  and manifest 
Lorentz invariance  
along with a global $SO(6)$  R-symmetry.  

We believe it is also possible to obtain a lattice action  and its  
continuum for the 
product gauge group $\prod_i U(k_i)$  where  $\sum_i k_i =k$ by expanding 
around a  point  in  the moduli space given by  
$\langle z^{\mu}_{\bfn } \rangle  = 
\frac{1}{\sqrt2 a} 1_k + ( \oplus_i c_i 1_{k_i})$. 
In the continuum limit, we  take the lattice spacing to zero by keeping 
$c_i$'s fixed, which  produce a $\prod_i U(k_i)$ gauge theory.  
It is desirable to know whether one can obtain  a 
lattice or matrix  regularization for orthogonal 
and symplectic gauge groups by either 
performing an orientifold projection on the  
 orbifold lattice or (deformed) type IIB matrix model (The symplectic type IIB 
is discussed in \cite{Itoyama:1998et}).  Such constructions may 
be useful in nonperturbative study of  electric-magnetic duality 
\cite{Montonen:1977sn}.

{\it Noncommutative continuum limit}

We believe it  is also  possible to obtain the noncommutative continuum 
$\CN=4$ theory  with the appropriate choices of the background flux matrix 
$\Phi_{\mu \nu}$, and hence appropriate zero action configurations
$U_{\mu}^{0}$,   for example the ones given in  \cite{Ambjorn:2000cs}.  
In order to achieve this limit, one needs to take $\theta'$ fixed  (hence  
$\Theta$ fixed) along with a fixed finite  volume.
In this case, there is a 
nonlocality of order $\sqrt {\Theta}$ in the star product of the
continuum theory. At
 distances longer then  $\sqrt {\Theta}$, 
the interactions  effectively shuts off because of the fast 
oscillation of the kernel. 
The action of the noncommutative $\CN=4$ SYM target theory can be obtained 
by replacing the product of fields in   \Eq{target4} by nonlocal 
continuum star product given in \Eq{contstar}. 

\section{ Collective excitations  of D(-1) branes as  D3 branes}
\label{sec:5}
The results obtained within this paper can be interpreted within the string
theory framework. In order to do that, we need to specify  the  
 world-volume theories of the corresponding    D-brane configurations.  
The type IIB matrix theory \Eq{typeIIB} 
describes the dynamics of a cluster of D(-1)-branes (which are 
sometimes referred as D-instantons).  These are points in ten dimensional 
Euclidean  space. In this sense  a Dp-brane is a p+1  (Euclidean) dimensional 
hypersurface.  
The  main result of this paper may be
interpreted as follows: Upon deforming the  type IIB matrix
theory,  a particular collective excitation of D(-1)-branes becomes the
classical vacuum state of the theory.  Essentially, the D(-1) branes, instead
of being at some random points on the moduli space of the undeformed theory,  
lie on the surface 
of a four dimensional torus $T^4$. 
 Analyzing  the fluctuations around 
this   D(-1)  background  yields   $\CN=4$ supersymmetric Yang-Mills theory, 
which is the  world-volume theory of a collection of 
D3-branes. Hence the D3-brane, which is  not present in the
fundamental description of the theory, may be considered as an emergent
brane.  In the rest of this section, we will describe these statement in a 
little bit more detail.  

The type IIB action \Eq{typeIIB} (or \Eq{act2}) 
 with gauge group $U(N_c)$ describes a collection of 
$N_c$  D(-1)-branes (see for example \cite{Polchinski:1998rr}).  
This  theory possess $\CQ=16$ supersymmetries and a
global $SO(10)$ symmetry. In the latter form of the action  \Eq{act2}, 
only 
the  $SU(5) \times U(1)$ subgroup of $SO(10)$ is manifest. 
The commutative moduli space is (locally) $\mathbb R^{10\Nc}$ 
(or equivalently  ${\mathbb C}^{5\Nc}$). The eigenvalues of these matrices  
($v_\alpha$ or $z^m$)
are identified as position of D(-1) branes.  In this framework, 
the  D-branes  are pointlike objects on  the moduli space. 

The deformed Type IIB  action  \Eq{deformed2}
 is  a supersymmetric deformation 
respectful to a global $U(1)^5$ symmetry and $\CQ=1$  supersymmetry.    
The deformation renders the moduli space (defined by $E$ and d-flatness
conditions given in \Eq{minima}) a noncommutative one,  and 
forces the eigenvalues of $z_m$ matrices (the D(-1) branes) 
to be uniformly distributed 
on a circle as in \ref{fig:moduli}.   However, as the matrices 
$z_1, \ldots  z_5$ are not simultaneously diagonalizable, the positions 
of the D(-1)  branes are indeterminate and they lie 
 on the surface of a  fuzzy four dimensional torus.   
The classical zero action configuration, given in \Eq{mod1} and  
\Eq{mod2}, is a collective excitation  of D(-1)-branes 
 and is 
responsible  for the emergence of higher dimensional branes out of the 
deformed Type IIB matrix theory. Also notice that,  the base space of the 
target theory is not the space that the eigenvalues reside, but the 
dual space obtained by Fourier transform  \Eq{realspace} 
in section \ref{sec:fluc}. The 
continuum limit is taken by keeping the size  of this dual space fixed.  
This is 
similar to the Seiberg-Witten limit where the open string metric 
and coupling constant   rather then the closed string parameters  are 
 kept fixed \cite{Seiberg:1999vs} 
(also see \cite{Adams:2001ne, Dorey:2003pp}).

It is also reasonable to ask which  supergravity field or type IIB string
theory field may be responsible for this excitation  of D(-1) branes.  
To answer  this question, let us first recall that 
the world-volume of a Dp-brane couples to a (p+1) form potential $A^{(p+1)}$, 
given in the form $\int_{Dp} A^{(p+1)}$,or explicitly as   
$\int A^{\mu_1 \ldots \mu_{p+1}} \frac{dx_{\mu_1}}{d \sigma_{i_1}} \ldots  
\frac{dx_{\mu_{p+1}}}{d \sigma_{i_{p+1}}} \epsilon^{i_1 \ldots i_{p+1}} 
d^{p+1} \sigma $, where  $x_{\mu}$ are world-sheet coordinates and the 
integration is over the world-sheet of Dp-brane.     
The  D(-1) branes, whose world-sheet is zero dimensional, 
 couples to a zero form. The  configuration we 
consider above carries $N_c$  units of D(-1)-brane charge. Moreover, 
the fact that the matrices $z_{m}$ do not commute with each other can be 
interpreted as  
the existence of higher dimensional branes in the system (see
\cite{Taylor:1997dy} for a review).
In our  case, the emergent higher dimensional branes are 
 $k$  D3-branes wrapped 
on a four dimensional spacetime hypertorus, where  $k$ is 
associated with the rank of  the gauge group 
$U(k)$ of the target theory. It is, therefore,  tempting to think that  
the deformations introduced  in \Eq{deformed2} on the matrix 
theory side corresponds to turning on the RR five-form self dual 
field  strength  $F^{(5)}$  on the type IIB string theory. The  $F^{(5)}$
arises from a four-form field $A^{(4)}$, which, as noted above, couples to  
D3-branes.  The emergence of  D3 branes can be
thought as  a special case of the dielectric brane  effect introduced 
by Myers, which is analogous to polarization of a neutral
atoms in an electric field \cite{Myers:1999ps}. Similarly, the collective 
state of D(-1) branes 
can be thought of as analogous to the 
 emergence of a dipole  upon the application of  $F^{(5)}$
field. However, the argument we have 
given here is incomplete  as it does
not specify the  geometry  or topology of the emergent object. 
For example, 
if we were to construct  the analog of  $\CN=1^{*}$ theory in four dimension 
  \cite{Polchinski:2000uf}, 
i.e,  a mass deformed matrix theory \cite{Andrews:2005cv},
 the emergent brane would be a (euclidean) 
D1 brane wrapped  on a two sphere $S^2$, whereas the deformation discussed in 
footnote 8 gives  a D1 wrapped on a two torus $T^2$.  Both of these 
can be thought as polarization of D(-1) branes.   The difference is similar to 
the emergence of dipole or quadrupole moments upon the application of 
appropriate electric fields.  Certainly, 
the string theory realization of the deformations    warrants further study. 

It is also possible to recover the known  membranes constructions  appearing 
in  Type IIB matrix theory  from the large $N_c$ limits of our formulation.   
Let us recall the zero action configurations $z_{\mu}= c_{\mu}U_{\mu}^{(0)}$
and  restrict our attention to the small fluctuations of the 
$U_{\mu}^{(0)}$,   and small $\Phi_{\mu \nu}$.  
Then   the conditions  describing the vacua 
$U_{\mu}^{(0)} U_{\nu}^{(0)}= e^{i \Phi_{\mu \nu}} U_{\nu}^{(0)}
U_{\mu}^{(0)}$  become   $[X_{\mu},   X_{\nu}]= -i\Phi_{\mu \nu}$, which
describes (locally)  a planar membrane, where $X_{\mu}$ are the 
fluctuations. 
The latter equation does not admit a solution 
in terms of finite matrices, and is usually considered to be associated with 
the existence of a D(p+2) in a system of Dp branes.  However,  notice that 
such a configuration is not stable within the Type IIB, 
but its decay is is suppressed in the large $N_c$  \cite{Aoki:1999vr}.   
The zero action configurations  of the deformed type IIB,  
on the other hand, admit solutions in terms of finite dimensional matrices, 
and, more
importantly,   
are stable (This point is also emphasized by Dorey \cite{Dorey:2003pp}).  
It is also worth mentioning that the analogs of 
a planar membrane, in the notation of
\Eq{deformed2},  is a  Fayet-Illiopoulis term,  given in the form of a d-term 
deformation  
$
\zeta \Tr   \int d \theta  \; \bfl  = - \zeta \, d 
$,
where $\zeta$ is the deformation parameter. This alters the d-flatness 
condition into 
$\sum_{m=1}^{5} [\mybar z_m, z^m ]- \zeta = 0  $
which only admits solution in terms of infinite dimensional matrices.  


\acknowledgments
I am  grateful for conversations about this work with 
Nima Arkani-Hamed, 
David B. Kaplan, Andreas Karch, Pavel Kovtun, Matt Strassler, Washington
Taylor, Larry Yaffe. I thank  Adam Martin for reading the manuscript and 
suggestions.  This work was supported by DOE grant  DE-FG02-91ER40676. 

\appendix
\section{Derivation of deformed type IIB action}
\label{sec:ap1}
In this appendix  we show that the dimensional reduction  of the lattice
action  \Eq{lattice}
 to one site model, 
by using the twisted boundary conditions \Eq{twisted},   gives the deformed 
type IIB action \Eq{deformed2}. We will first perform the 
reduction.  However, this will produce some  extraneous gauge rotation
matrices, which we will get rid of by appropriate field redefinitions.  

The first two term in action   \Eq{lattice} reduce to a form 
\beq 
\Tr\left [ -\frac{1}{2} {\bf \Lambda} {\partial}_{\theta} {\bf \Lambda}  
- {\bf \Lambda} [ (g_m^{\dagger} \mybar z_{m}),  
({\bf Z}_{m}g_m)] \right]_{\theta}
\eeq
which has no flux terms. This is  due to the fact that  
 the interaction of site fields with the link fields  does not surround an 
area. 

Next, we consider the interactions of fermi multiplets ${\bf \Xi}_{mn}$ with 
holomorphic  ${\bf E}^{mn}$ function.  First, notice that 
the   $\theta$ component of the ${\bf \Xi}_{mn}$ fermi 
multiplet  is a composite field,   and upon reduction it  becomes 
\beq
&& \mybar z_{m,\bfn+ {\bf \hat e_{n}} }  \mybar z_{n,\bfn}- 
\mybar  z_{n,\bfn+ {\bf \hat e_{m}} }  \mybar z_{m, \bfn} 
\rightarrow  (g_{n}\mybar z_{m} g_{n}^{\dagger})  \mybar z_{n}  -
 (g_{m} \mybar z_{n}g_{m}^{\dagger})  \mybar z_{m}
\cr
&&=  g_{n} g_{m} (g_{m}^{\dagger} \mybar z_{m}) (g_{n}^{\dagger}  
\mybar z_{n})-
g_{m}  g_{n} (g_{n}^{\dagger} \mybar z_{n})(g_{m}^{\dagger}  
\mybar z_{m})
= g_{m}  g_{n} \left( e^{-i \Phi_{mn}} (g_{m}^{\dagger} 
\mybar z_{m}) (g_{n}^{\dagger}  \mybar z_{n}) -
(g_{n}^{\dagger} \mybar z_{n})(g_{m}^{\dagger}  \mybar z_{m})  
\right) \qquad
\eeq
where we  used 't Hooft algebra. Hence  the third term in the 
action becomes      
\beq
\Tr \left[
{\bf \Xi}_{m n, \bfn}\, {\bf E}^{m n}_{\bfn}    \right]_\theta  \rightarrow 
\Tr \left[ (g_{n}^{\dagger} g_{m}^{\dagger}{\bf \Xi}_{mn}) 
 [  \;   (\bfz^{m} g_{m})
(\bfz^{n} g_{n})  - e^{i \Phi_{m n}} \; (\bfz^{n} g_{n})
(\bfz^{m} g_{m}) ]  \right]_\theta
\eeq
Even though we combined the two terms ${\bf \Xi}_{\mu \nu} {\bf E}^{\mu \nu} + 
2{\bf \Xi}_{\mu 5} {\bf E}^{\mu 5}  $ into 
${\bf \Xi}_{m n} {\bf E}^{m n} $, these are plaquettes of different types,
namely a 2-1-1 and 4-1-3 types 
(see fig.\ref{fig:plaquette}). 

We are done with the   $Q$-exact part of the lattice action. 
The $Q$-closed part  of action includes 
plaquettes of two types. One is 4-2-2 type, shown in 
fig.\ref{fig:plaquette} and the other is 3-1-2 type  shown in 
fig.\ref{fig:plaquette} and more explicitly, in  fig.\ref{fig:flux}.
The first type involves   
triangular loops composed of the 4-cell link field and two 2-cell fermionic 
links. The second type involves a 3-cell fermionic link, a 2-cell fermionic 
link and a bosonic 1-cell link field. Let us first consider the first type 
and its reduction. 
Using the action \Eq{lattice} and setting  $p=5$, we obtain 
\beq
&&  \frac{\sqrt 2 }{8}  \epsilon^{\mu \nu  \rho \sigma} {\bf \Xi}_{\mu \nu, 
\bf n}
(\mybar z_{5, \bf {n-  \hat e_5}}  {\bf \Xi}_{\rho \sigma , 
\bf {n +  \hat e_{\mu} + 
\hat e_{\nu} }} - 
{\bf \Xi}_{\rho \sigma, \bf {n -  \hat e_{\rho} - \hat e_{\sigma }} }
\mybar z_{5, \bf {n +  
\hat e_{\mu} + \hat e_{\nu }}}) \rightarrow  \cr
&&  \frac{\sqrt 2 }{8}  \epsilon^{\mu \nu  \rho \sigma} {\bf \Xi}_{\mu \nu}
( (g_5^{\dagger}\mybar z_{5}g_5)  (g_{\mu} g_{\nu} {\bf \Xi}_{\rho \sigma} 
g_{\nu}^{\dagger} g_{\mu}^{\dagger}) - 
(g_{\sigma}^{\dagger} g_{\rho}^{\dagger} {\bf \Xi}_{\rho \sigma} 
g_{\rho} g_{\sigma})
(g_{\mu} g_{\nu} \mybar z_{5} g_{\nu}^{\dagger} g_{\mu}^{\dagger})  
)
\cr 
&&  \frac{\sqrt 2 }{8}  \epsilon^{\mu \nu  \rho \sigma} 
( g_{\nu}^{\dagger} g_{\mu}^{\dagger} {\bf \Xi}_{\mu \nu})
\left( (g_5^{\dagger}\mybar z_{5})\, (g_5  g_{\mu} g_{\nu} g_{\rho}g_{\sigma})
\, (g_{\sigma}^{\dagger} g_{\rho}^{\dagger} {\bf \Xi}_{\rho \sigma}) 
 - 
(g_{\sigma}^{\dagger} g_{\rho}^{\dagger} {\bf \Xi}_{\rho \sigma} )
(g_{\rho} g_{\sigma}
g_{\mu} g_{\nu}g_{5})  (g_{5}^{\dagger} \mybar z_{5})  
\right) \qquad
\eqn{242}
\eeq
In this expression, since the combination   
$(g_5  g_{\mu} g_{\nu} g_{\rho}g_{\sigma}) \equiv R_{\mu \nu \rho \sigma}$ 
appears with  $\epsilon^{\mu \nu  \rho \sigma}$ tensor, with  the use of the 
definition  of $g_5$ \Eq{twisted}, we see that it 
is proportional to the identity matrix  and can be pulled out of the 
parenthesis. 
By using the 't Hooft's algebra \Eq{thooft} 
in the 
reordering of  the second term, 
we express \Eq{242} in a simpler form as  
\beq
\frac{\sqrt 2 }{8}  \epsilon^{\mu \nu  \rho \sigma}  R_{\mu \nu  \rho \sigma}
( g_{\nu}^{\dagger} g_{\mu}^{\dagger} {\bf \Xi}_{\mu \nu})
\left( (g_5^{\dagger}\mybar z_{5})\, 
\, (g_{\sigma}^{\dagger} g_{\rho}^{\dagger} {\bf \Xi}_{\rho \sigma}) 
 -  e^{i (\Phi_{\rho \mu} + \Phi_{\rho \nu} + 
\Phi_{\sigma \mu} +\Phi_{\sigma \nu} )} 
(g_{\sigma}^{\dagger} g_{\rho}^{\dagger} {\bf \Xi}_{\rho \sigma} )
(g_{5}^{\dagger} \mybar z_{5})  
\right) 
\eeq
where  the exponent is a  sum of the fluxes passing through all 
surfaces but $\mu \nu$  and $\rho \sigma$.

The 3-1-2 type plaquettes in $S_{closed}$ can be obtained by setting 
$p\neq 5$ in \Eq{lattice}.  
Following very similar steps to the reduction  
performed  above,  we obtain 
 \beq
&& 4 \frac{\sqrt 2 }{8}  \epsilon^{\mu \nu \rho \sigma
} {\bf \Xi}_{\mu \nu , \bf n}
(\mybar z_{\rho, \bf {n-  \hat e_{\rho}}}  {\bf \Xi}_{\sigma5, \bf {n +  \hat e_{\mu} + 
\hat e_{\nu} }} - 
{\bf \Xi}_{\sigma 5, \bf {n -  \hat e_{\sigma} - \hat e_5 }} \mybar z_{\rho, \bf {n +  
\hat e_{\mu} + \hat e_{\nu} }} )  \rightarrow \cr
&& \frac{\sqrt 2 }{2}  \epsilon^{\mu \nu \rho \sigma
} R_{\mu \nu \rho \sigma}
  (g_{\nu}^{\dagger}  g_{\mu}^{\dagger}{\bf \Xi}_{\mu \nu})
\left ( (g_{\rho}^{\dagger} \mybar z_{\rho}) \;  
(g_{5}^{\dagger}  g_{\sigma}^{\dagger}  {\bf \Xi}_{\sigma5})   
e^{i (\Phi_{\rho \mu} +\Phi_{\rho \nu})} 
 - 
 (g_{5}^{\dagger}  g_{\sigma}^{\dagger} {\bf \Xi}_{\sigma 5})  
(g_{\rho}^{\dagger} \mybar z_{\rho})  
  \right) 
\eeq

By doing a field redefinition, we can easily remove all  the gauge rotation 
matrices. An appropriate choice that we can read off from above
considerations is: 
\beq
g_{m}^{\dagger} \mybar z_{m} \rightarrow  \mybar z_{m}, \qquad 
\bfz^{m}g_m \rightarrow \bfz^{m}, \qquad 
 g_{n}^{\dagger} g_{m}^{\dagger} {\bf \Xi}_{mn}  \rightarrow {\bf \Xi}_{mn}
\eeq
The action of the deformed theory takes the simple form:
\beq
&&S_{d} = \frac{1}{g^2} \Tr   \left[ -\frac{1}{2} {\bf \Lambda} 
{\partial}
_{\theta} {\bf \Lambda}  - {\bf \Lambda} [\mybar z_m, {\bf Z}^m]  + \frac{1}{2}
{\bf \Xi}_{mn}({\bf Z}^m {\bf Z}^n -e^{i \Phi_{mn}}{ \bf Z}^n{\bf Z}^m)  
\right]_\theta +  \cr
&& \frac{\sqrt 2 }{2}  \epsilon^{\mu \nu  \rho \sigma}  
R_{\mu \nu  \rho \sigma}
 {\bf \Xi}_{\mu \nu}
\left[  
(\mybar z_{\rho} \;  
  {\bf \Xi}_{\sigma5}   
e^{i (\Phi_{\rho \mu} +\Phi_{\rho \nu})} 
 - 
 {\bf \Xi}_{\sigma 5}  
\mybar z_{\rho})  +
\frac{1}{4} 
(\mybar z_{5} \, \,  {\bf \Xi}_{\rho \sigma} 
 -  
 e^{i (\Phi_{\rho \mu} + \Phi_{\rho \nu} + 
\Phi_{\sigma \mu} +\Phi_{\sigma \nu} )}
{\bf \Xi}_{\rho \sigma} 
\mybar z_{5}  )
\right] \qquad 
\eqn{deformed1}
\eeq
where the usual commutators are altered to commutators involving 
fluxes. 

However, the above form of the action is not completely satisfactory. First, 
there are phases  $R_{\mu \nu \rho\sigma}$ in second line of  \Eq{deformed1}, 
and second,  
the redefined variables  are not perfectly 
antisymmetric. For example, the redefined   
${\bf \Xi}_{\mu \nu} \neq  - {\bf \Xi}_{\nu \mu}$,  but 
${\bf \Xi}_{\mu \nu} = - e^{-i\Phi_{\mu \nu}} {\bf \Xi}_{\nu \mu}$.  In
fact, our redefinition of the fields does not respect to the antisymmetry of 
the $p$-form fields (or it only respects up to a phase factor).   
 These two problems are indeed tied,  and can be cured by a 
better field redefinition which makes the antisymmetry properties manifest:
\beq
&&  e^{-i \Phi_{\nu \mu}/2 } g_{\nu}^{\dagger} g_{\mu}^{\dagger} {\bf \Xi}_{\mu
\nu}  \rightarrow {\bf \Xi}_{\mu\nu} \cr 
&& e^{-i ( \Phi_{\sigma \rho} +  \Phi_{ \sigma \nu } +  \Phi_{ \rho \nu}) 
/2 } g_{\sigma} g_{\rho}  g_{\nu} {\bf \Xi}^{\nu
   \rho \sigma}  \rightarrow {\bf \Xi}_{\nu \rho \sigma} \cr 
&&  e^{-i (  \Phi_{\sigma \rho} +  \Phi_{\sigma \nu} +  \Phi_{\sigma \mu}+
\Phi_{\rho \nu} +  \Phi_{\rho \mu} +  \Phi_{ \nu \mu}) /2 } 
  g_{\sigma}   g_{\rho} g_{\nu}   g_{\mu}
{\mybar z}^{\mu \nu \rho \sigma}
  \rightarrow {\mybar z}^{\mu\nu \rho \sigma} 
\eeq
By using the fact that the combination $
R_{\mu \nu \rho \sigma } \, 
e^{-i (\Phi_{ \mu \nu } + \Phi_{ \mu \rho } + 
\Phi_{\mu \sigma } +\Phi_{\nu \rho } + \Phi_{\nu \sigma } + 
\Phi_{\rho \sigma}  )/2}  $ is a constant phase,
 and some simple algebra,  we obtain the action \Eq{deformed2}. This action, 
as already stated, is a $\CQ=1$ supersymmetry, and  $U(1)^5$ global symmetry
preserving deformation of  the type IIB matrix theory.

\bibliography{nc}
\bibliographystyle{JHEP} 
\end{document}